\documentclass[letterpaper,conference,10pt]{IEEEtran}

\IEEEoverridecommandlockouts

\usepackage{amsmath}
\usepackage{amssymb}
\usepackage{amsfonts}
\usepackage{stmaryrd}
\usepackage{graphicx}
\usepackage{xcolor}
\usepackage{booktabs}
\usepackage{multirow}
\usepackage{makecell}
\usepackage{array}
\usepackage{pifont}
\usepackage{colortbl}
\usepackage{hhline}
\usepackage{bm}
\usepackage{threeparttable}
\usepackage{enumitem}
\usepackage{xspace}
\usepackage{cite}
\usepackage{url}
\usepackage{stfloats}

\usepackage{algorithm}
\usepackage{algorithmicx}
\usepackage{algpseudocode}
\usepackage[hidelinks]{hyperref}

\newcommand{\method}{OptiPrime}
\newcommand{\protocol}{OptiEncode}
\newcommand{\compression}{OptiComp}
\newcommand{\dataflow}{OptiFlow}
\newcommand{\restored}[1]{#1}

\newcommand{\revCC}[1]{#1}
\newcommand{\revA}[1]{#1}
\newcommand{\revB}[1]{#1}

\newcommand{\revD}[1]{#1}

\begin{document}

\title{\method: Optimizing Private Inference through Protocol--Hardware Co-design}

\author{%
\IEEEauthorblockN{Jiangrui Yu\textsuperscript{1},
Ye Yu\textsuperscript{1},
Si Chen\textsuperscript{2},
Chenqi Lin\textsuperscript{1},
Wenxuan Zeng\textsuperscript{1},
Junfeng Fan\textsuperscript{2},
Mingyu Gao\textsuperscript{3}, and
Meng Li\textsuperscript{1,*}}
\IEEEauthorblockA{\textsuperscript{1}Peking University, Beijing, China\quad
\textsuperscript{2}Open Security Research, Shenzhen, China\quad
\textsuperscript{3}Tsinghua University, Beijing, China}
\IEEEauthorblockA{Corresponding author\textsuperscript{*}}
\IEEEauthorblockA{\{jiangrui.yu, 2100012750, linchenqi\}@stu.pku.edu.cn,
\{si.chen, fan\}@osr-tech.com,\\
zwx.andy@outlook.com, gaomy@tsinghua.edu.cn, meng.li@pku.edu.cn}
}

\maketitle

\begin{abstract}

Private deep neural network (DNN) inference based on hybrid homomorphic encryption (HE) and multi-party computation (MPC) can protect user data with a formal guarantee, but at the cost of significant latency overhead due to HE. Customized HE accelerators have been proposed and have achieved orders-of-magnitude speedup for individual HE operations. However, when directly applying a commercial HE accelerator to state-of-the-art HE-MPC frameworks, we observe only limited end-to-end performance gain. This is because HE-MPC frameworks often require wireless transmission of input and output ciphertexts for each HE operation, leading to a severe network communication bottleneck.

To overcome this challenge, we introduce~\method, a protocol-hardware co-optimization framework for efficient private DNN inference. \method~features a novel HE protocol for convolutions that substantially reduces the number of transmitted output ciphertexts and mitigates the network communication bottleneck. Meanwhile, as the new protocol introduces complex computation for fewer output ciphertext, we observe new memory access challenges due to a high volume of weight plaintexts and intermediate ciphertexts. Hence, we further propose a lightweight compression system for the weight plaintexts, reducing memory traffic by 10$\times$, as well as a specialized dataflow to maximize on-chip data reuse of intermediate ciphertexts. Extensive experiments show that our framework outperforms the Cheetah baseline by at most $5.7\times$ on CPUs and $4.2\times$ with an accelerator.
Our code is available at \url{https://github.com/PKU-SEC-Lab/SEC-PPDL-LIB}.


\end{abstract}

\begin{IEEEkeywords}
Private inference, homomorphic encryption, multi-party computation, hardware acceleration, protocol--hardware co-design
\end{IEEEkeywords}

\section{Introduction}
\label{sec:intro}

The last decade has witnessed the rapid evolution of deep learning (DL) and its increasing adoption in privacy-sensitive applications, including medical diagnosis \cite{kononenko2001machine}, face recognition \cite{zhao2003face}, financial system \cite{kumar2020applicability}, etc. Privacy has therefore emerged as a major concern, leading to a growing demand for privacy-preserving DL (PPDL)
\cite{mireshghallah2020privacy,liu2020privacy,reagen2021cheetah,hao2022iron,xu2023falcon}.

\begin{figure}[!tb]
    \centering
    \includegraphics[width=\linewidth]{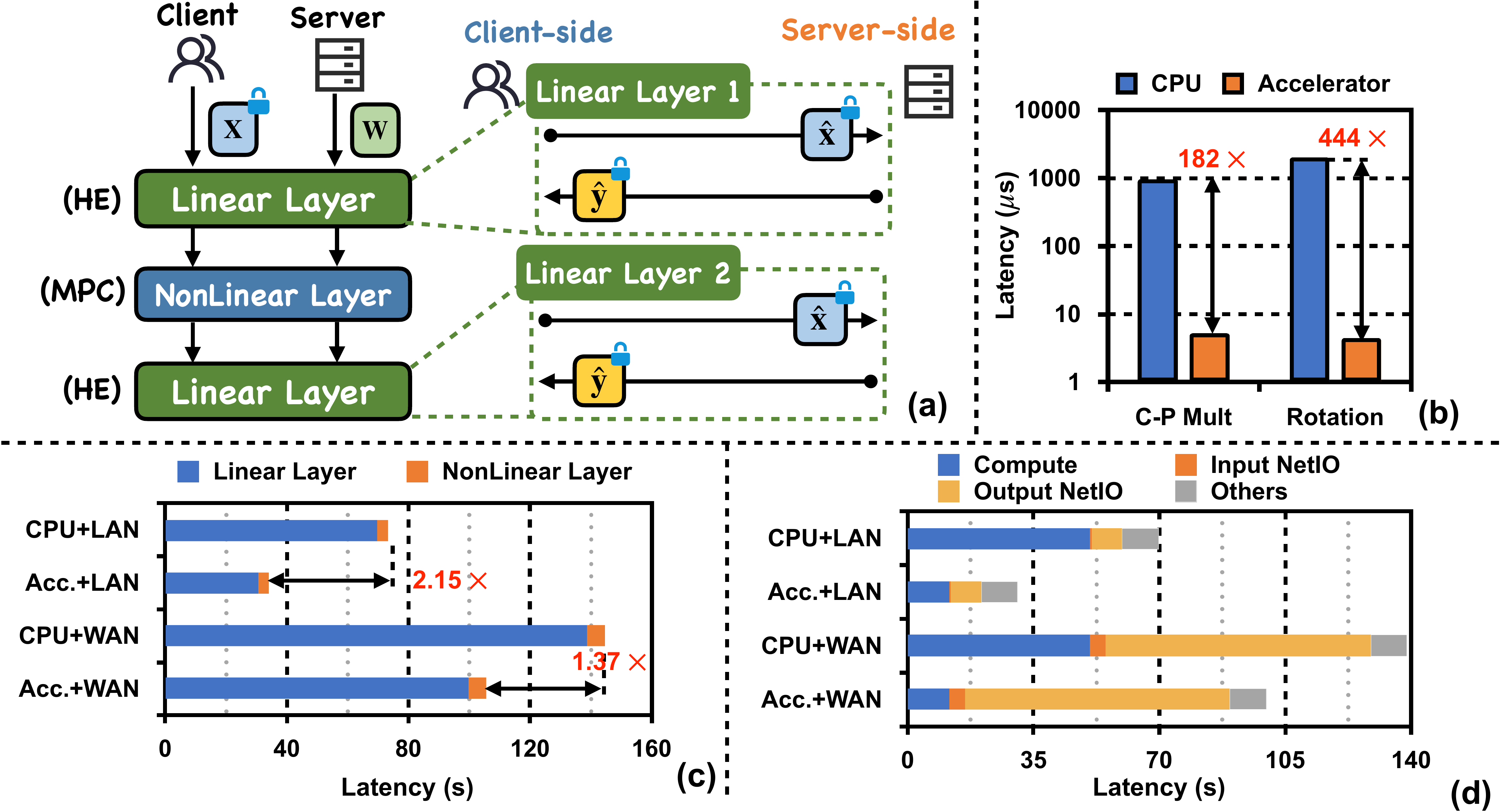}
    \caption{(a) Hybrid HE-MPC framework. (b) Latency reduction of HE operations with the hardware accelerator. (c) Latency breakdown of ImageNet-scale ResNet50 under different network and computation conditions. "Acc." means the accelerator. (d) Latency breakdown of the Linear layer of ResNet50. "Compute" represents the HE computation. "Input NetIO" and "Output NetIO" represent the server-client network transmission of the input ciphertexts and the output ciphertexts, respectively. "Others" represents other CPU overhead, like ciphertext decryption.}
    \label{fig: latency}
\end{figure}

PPDL frameworks based on hybrid Homomorphic Encryption (HE) and Multi-Party Computation (MPC) have recently been proposed and have attracted a lot of attention \cite{rathee2020cryptflow2,rathee2021sirnn,mohassel2017secureml,juvekar_gazelle_2018, Mishra2020delphi,hao2022iron,pang2023bolt,lu2023bumblebee,xu2024privcirnet,xu2023falcon,singh2024hyena,huang2022cheetah,yu2024flexhe,li2024nimbus,hyunjun2024apint,balla2023heliks,garimella2023characterizing,xu2025blb,zhang2025fenix}. As shown in Figure~\ref{fig: latency} (a), an HE-MPC framework often involves two parties, namely the server and the client, which own private deep neural networks (DNNs) and input data, respectively. The two parties jointly execute a series of protocols, including HE for linear operations (e.g., convolutions) and MPC for nonlinear functions (e.g., ReLU), so that the final results can be computed while the privacy of both input data and DNN parameters can be preserved \cite{mohassel2017secureml,rathee2020cryptflow2,rathee2021sirnn}.

An alternative approach for PPDL is to leverage end-to-end fully HE (FHE) \cite{ju2024neujeans,lee2022low,reagen2021cheetah,stoian2023deep,park2022aespa,bachrachv48cryptonets, Liu2017ObliviousNN,chet2019roshan,Aharoni2023helayers,ebel2025orion,park2024powerformer,ran2022cryptogcn,ran2023penguin}. It computes all DNN operations based on HE and avoids the interaction between the server and the client. However, it often requires extensive approximation for nonlinear activation functions and expensive bootstrapping operations, which may suffer from accuracy bottlenecks \cite{lou2020falcon}. \textbf{Therefore, in this paper, we focus on the HE-MPC framework.}



\begin{figure*}[!tb]
    \centering
    \includegraphics[width=\linewidth]{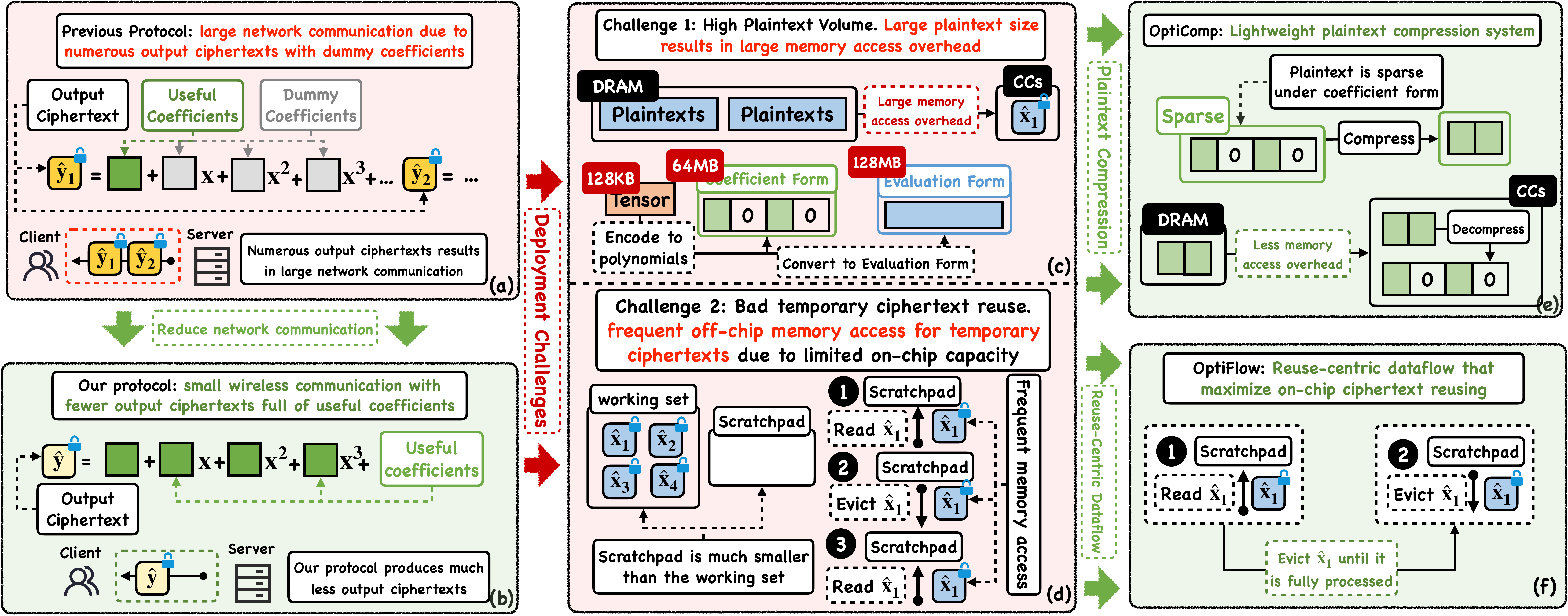}
    \caption{Overview of \method, including the limitations of previous protocol and our protocol, the challenges to deploy this protocol, and our solutions. }
    \label{fig: overview}
\end{figure*}

The hybrid HE-MPC framework often incurs high latency primarily due to the costly HE operations \cite{juvekar_gazelle_2018, huang2022cheetah}. To speed up HE operations, numerous HE acceleration schemes, including ASICs \cite{nikola2021f1,Kim2022bts,jongmin22ark,nikola2022craterlake,jongmin23sharp,deng2024trinity,fast2025fan,zhang2025flash,ebel2024osiris,zhou2024ufc, Prasetiyo2024Morphling, Agrawal2024heap}, FPGAs \cite{Sadegh2020heax,agrawal2022fab,yang_poseidon_2023,roy2019heat,2021hexlfpga,yang2025hydra}, and GPU libraries \cite{2018cufhe,dai2015cuhe,fan2022tensorfhe,jung2021over100x,kaust2023gme,fan2025warpdrive,jiaodian2025neo}, have been designed. However, when we apply a commercial FPGA-based accelerator \cite{lattisense} to the state-of-the-art (SOTA) Cheetah protocol \cite{huang2022cheetah}, only a modest \textbf{1.37$\times$} end-to-end latency reduction is achieved as shown in Figure \ref{fig: latency} (c), in stark contrast to its significant speedups of individual HE operations (Figure \ref{fig: latency} (b)). Further performance breakdown in Figure \ref{fig: latency} (d) reveals a critical insight: while the HE computation time is significantly reduced, the overall benefit is negated by the massive network communication overhead inherent to the hybrid HE-MPC framework.


As shown in Figure~\ref{fig: latency} (a), communication in the HE-MPC framework arises from two sources: (1) transmitting input and output ciphertexts for each HE-based linear layer, and (2) MPC protocols for nonlinear functions. Figure~\ref{fig: latency} (c) and (d) indicate that linear layers dominate the overall latency in Cheetah, with output ciphertext transmission as the primary bottleneck - particularly when HE accelerators are used. We observe that the inefficiency stems from Cheetah’s convolution protocol, which generates numerous ciphertexts because each ciphertext encodes only a small number of useful elements. Therefore, \textbf{to address this bottleneck, we propose \protocol, a new convolution protocol that significantly reduces the number of output ciphertexts (as in Figure~\ref{fig: overview} (b)).} While this lowers the communication overhead, it introduces more complex HE computation, leading to two additional memory access challenges.

\textbf{Challenge 1: High Plaintext Memory Volume.}
As shown in Figure~\ref{fig: overview} (c), \protocol~substantially increases memory demand for weight plaintexts. For HE operations, weights are encoded into polynomials, which are highly sparse with over 99\% of coefficients padded with zero. Furthermore, existing protocols often convert these polynomials from the coefficient form to the evaluation form to reduce runtime computation, further inflating their sizes \cite{sealcrypto, badawi22openfhe}. Consequently, memory requirements for weight plaintext polynomials increase by over 1000$\times$, creating a prohibitive bottleneck in weight plaintext fetching.



\textbf{Solution 1: Lightweight Plaintext Compression (\compression).}
To reduce memory capacity and bandwidth demands, we introduce \compression, a lightweight compression system (Figure~\ref{fig: overview}(e)). \compression~exploits the sparsity of coefficient-form polynomials by compressing them according to their sparse patterns. At runtime, compressed plaintexts are decompressed and converted to the evaluation form on the fly. Although this adds minor computation overhead, the substantial memory savings yield significant performance gains.



\textbf{Challenge 2: Bad temporary ciphertext reuse.}
Unlike Cheetah, \protocol~involves complex HE computations to reduce the number of output ciphertexts. As a result, many more intermediate ciphertexts are generated, whose working set size quickly exceeds the accelerator on-chip memory capacity (Figure~\ref{fig: overview}(d)) and results in frequent off-chip memory accesses for temporary data.

\textbf{Solution 2: Reuse-Centric Dataflow (\dataflow).}
To resolve this, we propose \dataflow, a specialized dataflow that reorders the computation to maximize data reuse (Figure~\ref{fig: overview}(f)). By carefully scheduling the HE operations, \dataflow~ ensures that subsets of the ciphertext working set are fully processed within the on-chip scratchpad before being evicted, thus minimizing costly off-chip memory traffic.

\revA{Extensive experiments show that our framework outperforms the Cheetah baseline by up to $5.7\times$ on CPUs and $4.2\times$ with an accelerator. On end-to-end tasks, it reduces the inference latency for ResNet-18 and ResNet-50 on ImageNet to just 2.9 seconds and 14.6 seconds, respectively.}








\section{Background}
\label{sec:background}

\subsection{Homomorphic Encryption and Encodings}

HE allows one party to perform computation, e.g., addition and multiplication, on encrypted data without decryption. We mainly focus on the Leveled HE (LHE) scheme based on ring learning with error (RLWE), i.e., BFV \cite{fan2012fv}, to compute linear layers. BFV computes on polynomials, and the main HE parameters include the polynomial degree $N$, the plaintext modulus $t$, and the ciphertext modulus $q$. As HE operates over 1-dimensional polynomials and DNN computes over high-dimensional tensors, mapping from tensors to polynomials, denoted as encoding, is important and directly determines the computation efficiency.

There are two major encoding schemes: \textbf{coefficient encoding} \cite{hao2022iron, lu2023bumblebee, huang2022cheetah,li2024nimbus, park2024powerformer} and SIMD encoding \cite{juvekar_gazelle_2018,Mishra2020delphi, pang2023bolt}. SIMD encoding enables element-wise addition and multiplication on encrypted vectors. However, it imposes a strict requirement that the plaintext modulus is of the form $2kN+1$, where $N$ is the polynomial degree and $k$ is a positive integer. This restriction can degrade the performance of MPC \cite{huang2022cheetah} and is often not preferred in HE-MPC frameworks.

\begin{figure}[!tb]
    \centering
    \includegraphics[width=1.0\linewidth]{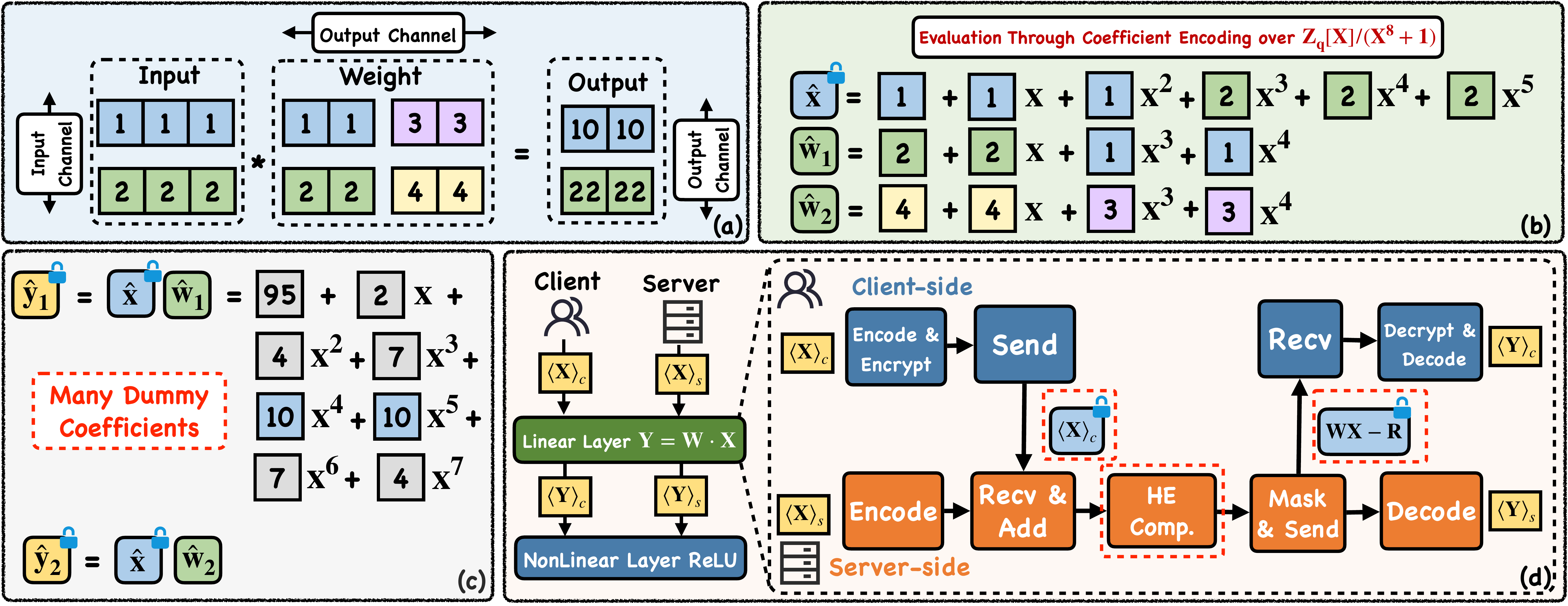}
    \caption{(a) An example of a convolution operation. (b) An example of coefficient encoding. (c) The output of the coefficient encoding. Two correct results generated are colored in blue, while the rest are dummy ones. (d) Overview of the Hybrid HE-MPC framework.}
    \label{fig: 3-framework}
\end{figure}

In contrast, coefficient encoding places elements directly in polynomial coefficients. For example, as shown in Figure \ref{fig: 3-framework} (b), the Cheetah protocol encodes input and weight tensor along their width dimension. After homomorphic computation, the correct results appear in specific coefficients of the output ciphertext, as illustrated in Figure \ref{fig: 3-framework} (c). It is inherently MPC-friendly and incurs lower communication cost \cite{huang2022cheetah}. Furthermore, it typically requires fewer HE operations for convolutions, as polynomial multiplication naturally implements convolution, as shown in Table~\ref{tab: method comparison}. \textbf{Therefore, in this paper, we mainly focus on the coefficient encoding.} 

\revA{A major limitation of coefficient encoding is the prevalence of ``dummy coefficients,'' which reduces encoding density and increases the number of output ciphertexts. As illustrated in Figure~\ref{fig: 3-framework}(c), only two coefficients colored in blue carry useful data, while the rest are dummy. Since polynomial multiplication convolves all coefficients, these dummy values cannot be eliminated with simple masking as in SIMD encoding. Consequently, although each ciphertext can hold eight coefficients in the example, four outputs are inefficiently distributed across two ciphertexts, leading to substantial communication overhead. \textbf{This highlights the need for a new protocol that produces fewer and more densely encoded ciphertexts.}}



\subsection{Hybrid HE-MPC Framework}

As illustrated in Figure \ref{fig: 3-framework} (d), the linear layer begins with the client and server each holding an ``additive share'' $\langle\mathbf{X}\rangle_c$ and $\langle\mathbf{X}\rangle_s$ of the input activation tensor $\mathbf{X}$, where $\mathbf{X}=\langle\mathbf{X}\rangle_c+\langle\mathbf{X}\rangle_s \mod{t}$, and t is the plaintext modulus.

First, the client encodes and encrypts its shares into $\left\llbracket\langle\mathbf{X}\rangle_c \right\rrbracket$ and sends it to the server. The server then locally adds to its share to recover the encrypted input: $\left\llbracket \langle\mathbf{X}\rangle_c \right\rrbracket + \langle\mathbf{X}\rangle_s = \left\llbracket \mathbf{X} \right\rrbracket$. Next, the server homomorphically evaluates the linear layer by computing $ \mathbf{W} \cdot \left\llbracket  \mathbf{X} \right\rrbracket - \mathbf{R} = \left\llbracket  \mathbf{W} \mathbf{X}  - \mathbf{R} \right\rrbracket $, where $\mathbf{R}$ is randomly generated to mask $\mathbf{Y}=\mathbf{X}\mathbf{W}$ and keeps as the server's share $\langle\mathbf{Y}\rangle_s$. This result is sent back to the client, who decrypts it to obtain its output share, $\langle\mathbf{Y}\rangle_c$. These shares of the output $\mathbf{Y}$ then serve as inputs for the subsequent non-linear layer, which is collaboratively evaluated using an MPC protocol like the accurate ReLU from CrypTFlow2~\cite{rathee2020cryptflow2}. The primary bottleneck of this framework lies in the HE-based linear layer (detailed in section \ref{sec: efficient conv}), which suffers from \textbf{server-side HE computation overhead and the back-and-forth network communication overhead of secret shares.} 

\begin{figure}[!tb]
    \centering
    \includegraphics[width=\linewidth]{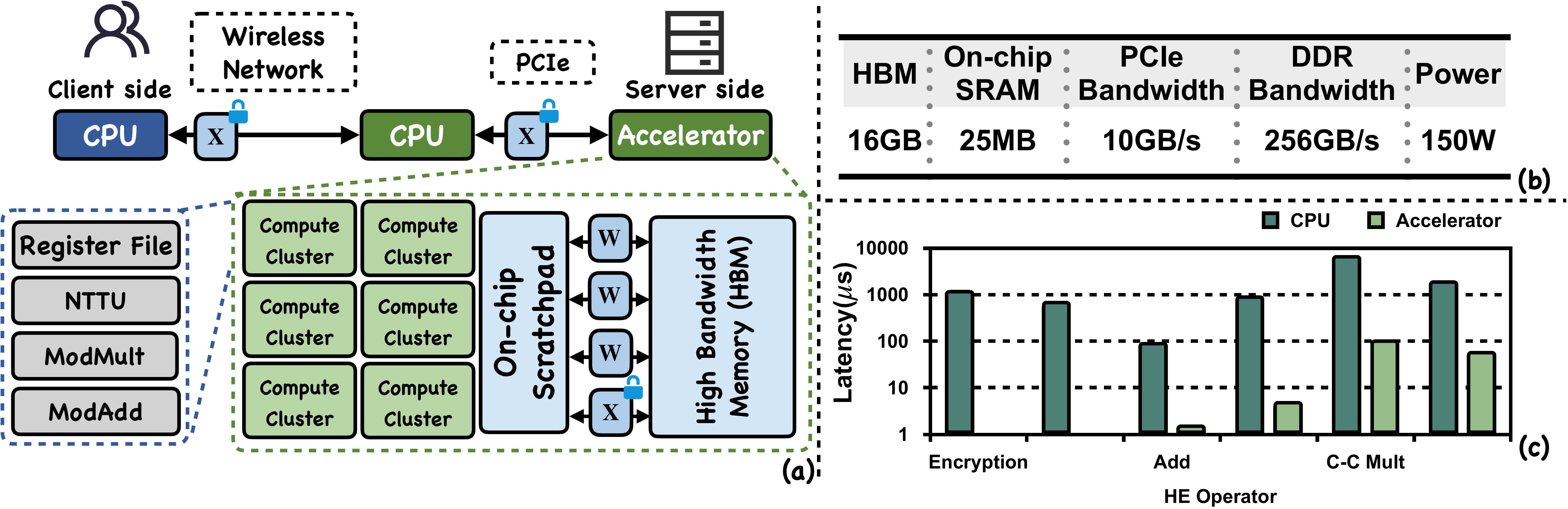}
    \caption{\revB{Latency of HE computation operations on the accelerator and CPU for parameters \mbox{$N=8192$} and \mbox{$\log q \approx 64$}.}}
    \label{fig: 3-cost}
\end{figure}

Figure \ref{fig: 3-cost}(a) presents the hardware system for our HE-MPC framework. Both the client and server are equipped with a CPU and communicate over a wireless network. We augment the server with an FPGA-based HE accelerator \cite{lattisense}, the configuration of which is detailed in Figure \ref{fig: 3-cost}(b).

\revA{This accelerator has an architecture similar to previous works such as F1~\cite{nikola2021f1} and ARK~\cite{jongmin22ark}, consisting of High Bandwidth Memory (HBM), an on-chip scratchpad, and multiple compute clusters (CCs). Each cluster contains specialized units for key HE operations, including NTT (NTTU), automorphism (AutoU), modular multiplication (ModMultU), and modular addition (ModAddU). We omit further implementation details of the accelerator because our contributions are hardware-agnostic and applicable to most existing HE accelerators, such as F1 and ARK.} The per-operation latency is shown in Figure~\ref{fig: 3-cost}(c). \revB{During inference, each ciphertext is first transmitted from the client to the server over the network, and then forwarded from the server CPU to the accelerator over PCIe. During the HE computation, the accelerator repeatedly loads ciphertexts and plaintexts from the HBM. Consequently, the network transmission and the HBM accesses constitute the two dominant sources of overhead.}

\subsection{Threat Model}

\revA{\method~adopts the standard two-party honest-but-curious model of prior hybrid HE--MPC frameworks~\cite{huang2022cheetah,rathee2020cryptflow2,hao2022iron,lu2023bumblebee,pang2023bolt}, in which each party follows the prescribed protocol honestly but may try to infer additional information about the other party's private inputs. The client holds a private input, the server holds private model weights, and the network architecture and tensor dimensions are public. BFV semantic security protects the client's encrypted share, the model weights never leave the server, and intermediate values are exchanged only in encrypted or secret-shared form. \protocol~changes only the public coefficient layout, whereas \compression~and \dataflow~are server-local. None introduces secret-dependent control flow, new messages, or additional interaction rounds. Thus, by standard composition, \method~retains the privacy guarantees of the underlying HE--MPC framework: the client learns only the prescribed inference output, while the input and model remain private.}


\section{\protocol: a Communication-Efficient Convolution Protocol}
\label{sec: efficient conv}

\subsection{Limitations of previous protocol}

\begin{figure}[!tb]
    \centering
    \includegraphics[width=\linewidth]{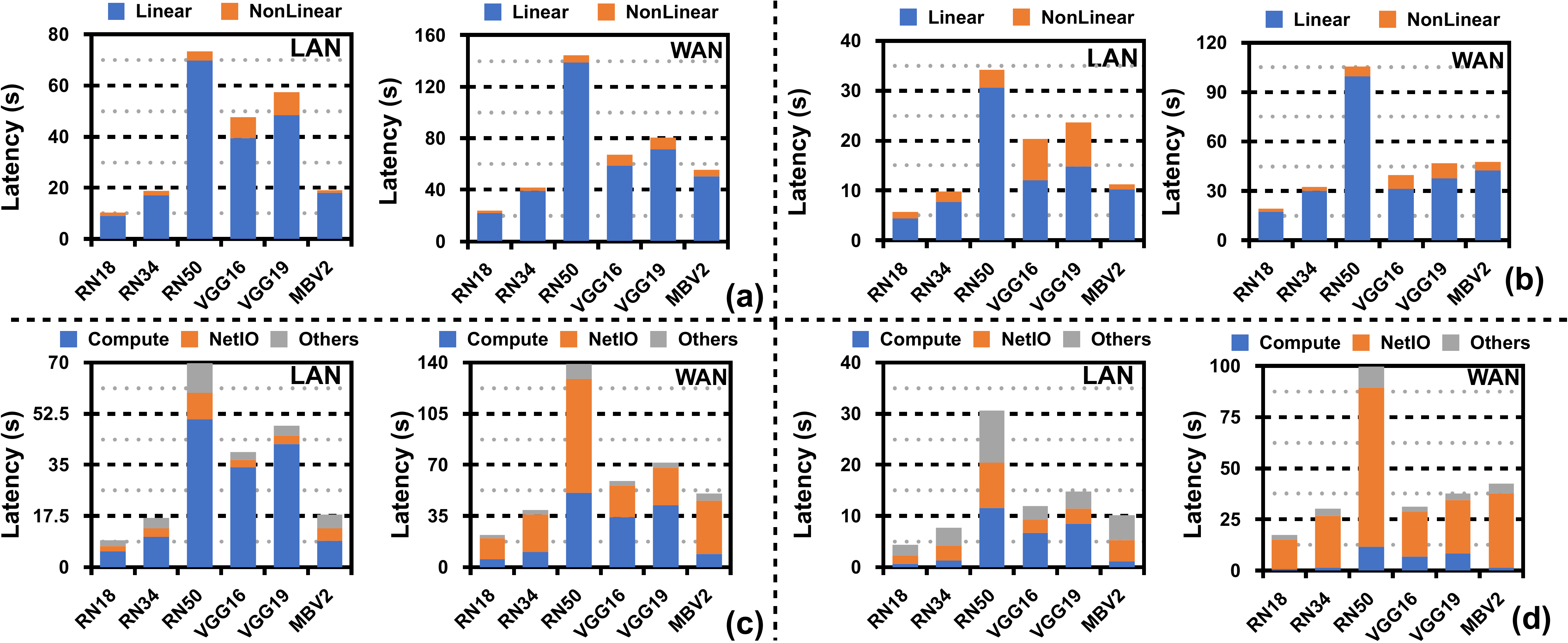}
    \caption{(a) End-to-end latency breakdown of CPU baseline. (b) End-to-end latency breakdown of the accelerator. (c) Linear layer latency breakdown of the CPU baseline. (d) Linear layer latency breakdown of the accelerator.} 
    
    \label{fig: 4-breakdown}
\end{figure}

We begin by profiling the current SOTA coefficient encoding-based convolution protocol, Cheetah \cite{huang2022cheetah}. As illustrated in Figure \ref{fig: 4-breakdown} (a) and (b), \textbf{the HE-based linear layer is the bottleneck with and without a server-side hardware accelerator}, which is consistent with previous work \cite{huang2022cheetah,singh2024hyena,li2024nimbus,juvekar_gazelle_2018}. This is because non-linear layers are less intensive and can be further optimized with accelerators \cite{tan2021cryptgpu,xing2022ppmlac,xiaolin2025POTA}. In contrast, the linear layers suffer from HE's high cost. \textbf{Therefore, in this work, we focus on optimizing the HE part (i.e., linear layer computation). }

However, as shown in Figure \ref{fig: 4-breakdown} (c) and (d), further latency breakdown of linear layers shows that after applying hardware acceleration, the bottleneck shifts from the HE computation to wireless network communication, leading to \textbf{limited overall speedup}. This excessive communication is a direct consequence of Cheetah's encoding, which generates numerous output ciphertexts filled with dummy coefficients. While SIMD-based protocols like Hyena \cite{singh2024hyena}, Orion \cite{ebel2025orion}, and Gazelle \cite{juvekar_gazelle_2018} also produce compact outputs, they incur more costly HE operations and MPC-incompatible prime moduli, as shown in Table~\ref{tab: method comparison}.

To address this, we propose \protocol, a communication-efficient protocol with two key components. First, a channel encoding encodes the input and weight tensor along the input channel dimension, which enables the use of automorphisms (i.e., rotations in SIMD encoding) to eliminate dummy coefficients \cite{lu2023bumblebee, chen2021ringconv}. Second, to reduce the cost of these automorphisms, we introduce a computational reformulation that enables the baby-step-giant-step (BSGS) algorithm.
Compared to Cheetah, our protocol reduces \textbf{output ciphertexts by at most $\mathbf{128\times}$}. With further reformulation optimization, our protocol reduces \textbf{automorphisms by at most $\mathbf{128\times}$}.


\subsection{Channel Encoding}

\begin{figure*}[!tb]
    \centering
    \includegraphics[width=\linewidth]{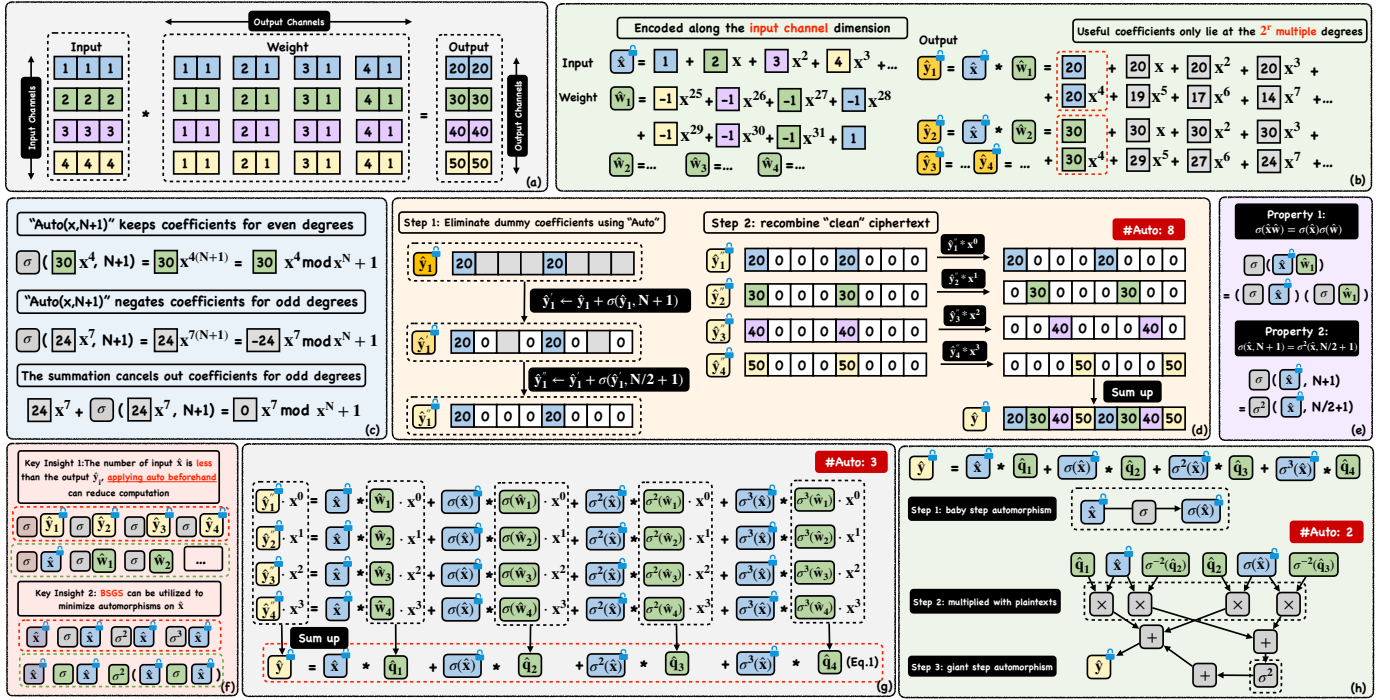}

\caption{\revB{(a) A toy example of the convolution operation. (b) An example of channel encoding. (c) The automorphism \mbox{$\sigma(\cdot, N+1)$}. (d) Eliminating dummy coefficients with automorphisms. (e) The key properties enabling the reformulation. (f) The key insights that guide the reformulation. (g) An example where all intermediate results (e.g., \mbox{$\hat{y}_i, \hat{y}'_i, \hat{y}''_i$}) are replaced with \mbox{$\hat{x}$} and \mbox{$\hat{w}$}, and unifying automorphisms under the base \mbox{$\sigma(\cdot, N/2+1)$}. (h) The proposed computation optimized with the BSGS technique.}}
    
    \label{fig: 4-encoding}
\end{figure*}

We consider the convolution $Y = X * W$ between a ciphertext tensor $X \in \mathbb{R}^{C_i \times H \times W}$ and a plaintext weight $W \in \mathbb{R}^{C_o \times C_i \times h \times w}$, which results in an output $Y \in \mathbb{R}^{C_o \times H \times W}$.
Throughout this section, we will use the configuration shown in Figure~\ref{fig: 4-encoding}~(a), where $C_o=C_i=4$, $H=h=1$, $W=3$, and $w=2$, as a running example for illustration.
To leverage automorphisms to clean up dummy coefficients (as detailed subsequently), the useful terms in the output polynomial must occupy degrees that are multiples of a power of two (i.e., degrees $ k\cdot2^r$). For instance, if the useful coefficients are at degrees $\{x^0, x^4, x^8, \dots\}$, automorphisms can eliminate the intermediate dummy coefficients (at $\{x^1, x^2, x^3, \dots\}$).

To satisfy this constraint, we design a specific encoding method, depicted in Figure~\ref{fig: 4-encoding}(b). \textbf{The key insight} is to encode the tensors $X$ and $W$ along the input channel dimension, $C_i$. This strategy ensures that the convolution directly yields the useful coefficients at the degrees: $\{x^0, x^{C_i}, x^{2C_i}, \dots\}$. If $C_i$ is a power of two (as in our example, where $C_i=4$), the useful coefficients are correctly positioned at the desired degrees $\{x^0,x^4,\dots\}$. We now formally define our encoding schemes, $\pi_{\text{conv}}^{x}$ and $\pi_{\text{conv}}^{w}$:
\begin{flalign*}
& \hat{x}=\pi_{\text{conv}}^{x}(\mathbf{X}) \text{ and } \hat{w} = \pi_{\text{conv}}^{w}(\mathbf{W}) \text{ such that} && \\
& \hat{x}[iWC_i+jC_i+c]=\mathbf{X}[c,i,j] && \\
& \hat{w}[0]=\mathbf{W}[C_o-1,C_i-1,H-1,W-1] && \\
& \begin{aligned}[t] 
    \hat{w}[N - C_i(c'HW + cW + j) - c] & \\ 
    \qquad = -\mathbf{W}[c',c,i,j] & 
  \end{aligned} 
  && (\text{otherwise}) 
\end{flalign*}

Figure~\ref{fig: 4-encoding}~(b) illustrates this encoding scheme. For the input tensor $X$, the first elements across all four channels (i.e., $\{1, 2, 3, 4\}$) are mapped to the first four coefficients of $\hat{x}$, followed by the second and the third elements of all input channels. For the weight tensor $W$, we adhere to the same channel-first encoding but arrange the elements in reverse order. Consequently, the polynomial multiplication $\hat{y}_1=\hat{x} \cdot \hat{w}_1$ yields valid convolution results at degrees that are multiples of $4$ (specifically, coefficients of $x^0$ and $x^4$), while intermediate terms (e.g., coefficients of $x^1, x^2, \dots$) contain dummy values. To accommodate tensors exceeding the capacity of a single polynomial of degree $N$, we partition the tensor along the input channel dimension into smaller subtensors and apply this encoding process to each partition.

With this encoding, we now demonstrate how to utilize automorphism operations to eliminate dummy coefficients. Formally, given a polynomial $\hat{y}=\sum_i c_i X^i$, the automorphism $\sigma(\hat{y},k)$ maps each term $c_i X^i$ to $c_i X^{ik} \pmod{X^N+1}$. By selecting specific indices for $k$, we can preserve certain coefficients while inverting the signs of others.

For instance, applying the automorphism with $k=N+1$ yields $\sum c_i X^{i(N+1)}$. Since $X^N \equiv -1 \pmod{X^N+1}$, this operation results in:
\begin{multline*}
        \sigma(\hat{y}, N+1) = \sum c_i X^i (X^N)^i = \sum c_i X^i (-1)^i \\
    = c_0 - \textcolor{red}{c_1}X + c_2X^2 - \textcolor{red}{c_3}X^3 + \dots \mod{X^N+1}
\end{multline*}

As illustrated in Figure~\ref{fig: 4-encoding}~(c), this operation preserves the coefficients of even-degree terms (e.g., $x^0, x^4$) while negating the coefficients of odd-degree terms (e.g., $x^1, x^3$). Therefore, computing the sum $\hat{y} + \sigma(\hat{y}, N+1)$ cancels the odd-degree terms and doubles the even-degree terms:
\[
\hat{y} + \sigma(\hat{y}, N+1) = 2c_0 + \textcolor{red}{0}X + 2c_2X^2 + \textcolor{red}{0}X^3 + \dots \mod{X^N+1}
\]
Subsequently, we apply the automorphism $k=N/2+1$. This operation targets terms with degrees that are odd multiples of two (i.e., $\{x^2, x^6, \dots\}$):
\[
\sigma(\hat{y}, N/2+1) = 2c_0 - 2c_2X^2 + 2c_4X^4 - \dots \mod{X^N+1}
\]
By iteratively applying this logic using the recurrence:
\[
\hat{y} \leftarrow \hat{y} + \sigma(\hat{y}, N/2^j+1) \quad \text{for } j=0, 1, \dots, r-1
\]
we effectively isolate the coefficients at positions that are multiples of $2^r$, and eliminate all remaining dummy coefficients.

Therefore, as shown in Figure~\ref{fig: 4-encoding}~(d), starting with the polynomial $\hat{y}_1$, we first compute $\hat{y}^{'}_1=\hat{y}_1 + \sigma(\hat{y}_1, N+1)$. This operation effectively eliminates dummy coefficients corresponding to odd degrees (e.g., $x^1, x^3, \dots$). Subsequently, we compute the next stage of the accumulation using $\hat{y}^{''}_1=\hat{y}^{'}_1+\sigma(\hat{y}^{'}_1, N/2+1)$, which removes coefficients at degrees such as $x^2$ and $x^6$. Ultimately, only the valid coefficients at degrees $x^0$ and $x^4$ are preserved, while coefficients at all other degrees are zeroed out.

The outputs are then recombined by first multiplying a plaintext polynomial $x^i$ to rotate and align the coefficients, and then sum together (Figure \ref{fig: 4-encoding} (d)). The preceding process shows that an automorphism's ability to eliminate dummy coefficients relies on a specific ciphertext encoding. Our proposed encoding meets this requirement, unlike Cheetah's output where useful coefficients are clustered together (Figure \ref{fig: 3-framework} (c)). 

Table \ref{tab: method comparison} provides a detailed comparison with other protocols, showing that our protocol produces fewer output ciphertexts but requires a large number of automorphisms on ciphertexts, potentially increasing computation time.

\subsection{Reducing Automorphisms: a BSGS Approach}

We now present a more efficient method for reducing automorphisms. This approach achieves a greater than $128\times$ reduction in automorphisms.

Our method leverages two properties illustrated in Figure \ref{fig: 4-encoding} (e): \underline{1)} automorphisms on the result $\hat{y}=\hat{x}\hat{w}$ can be applied beforehand on the inputs $\hat{x}$ and $\hat{w}$, and then perform the multiplication, which is $\sigma(\hat{y})=\sigma(\hat{x})\sigma(\hat{w})$.  \underline{2)} applying $\sigma(\cdot,N+1)$ to any $\hat{x}$ is equivalent to applying $\sigma(\cdot,N/2+1)$ twice. This relationship, $\sigma(\cdot,N+1)=\sigma^2(\cdot,N/2+1)$, holds because:
\[
\sigma^{2}_{N/2+1} (a_ix^i) = a_ix^{(\frac{N}{2}+1)^2i} = a_i x^{\frac{N^2i}{4}+Ni+i} = a_ix^{(N+1)i}
\]
where $x^{\frac{N^2i}{4}} = (x^N)^{\frac{Ni}{4}} \equiv (-1)^{\frac{Ni}{4}} \equiv 1 \pmod{X^N+1}$.
In general, the set of automorphisms $\{\sigma(\cdot,N+1),\sigma(\cdot,N/2+1),\dots\}$ can all be expressed as powers of a single base automorphism as $\{\sigma^{2^m},\sigma^{2^{m-1}},\dots,\sigma\}$, where $\sigma \equiv \sigma(\cdot, N/2^m+1)$. 
In the context of our running example, the required set of automorphisms is $\{\sigma(\cdot, N+1), \sigma(\cdot, N/2+1)\}$. By defining the base automorphism as $\sigma = \sigma(\cdot, N/2+1)$, this set can be represented as $\{\sigma^2, \sigma\}$.

\begin{table}[!tb]
  \centering
  \setlength{\arrayrulewidth}{1pt}
  \setlength{\doublerulesep}{1.5pt}
  \caption{Comparison with prior works. The concrete numbers are based on \revB{one layer of} the Imagenet-Resnet18 with $C_o= 256, C_i=256, H=14, W=14, N=8192.$ "$f=h\times w$" is the number of filter elements.}
  \resizebox{\linewidth}{!}{
    \begin{tabular}{ |c||c|c|c| }
      \hline
      \cellcolor{gray!25}{\textbf{Coeff.}} & \cellcolor{gray!25}{\textbf{Cheetah\cite{huang2022cheetah}}} & \cellcolor{gray!25}{\textbf{Ours w/o BSGS}} & \cellcolor{gray!25}{\textbf{Ours w/ BSGS}} \\
      \hline\hline
      \cellcolor{gray!25}{\#CPMult} & \makecell{$O(HWC_{i}C_{o}/N)$ \\ 2048} & \makecell{$O(HWC_{i}C_{o}/N)$ \\ 2048} & \makecell{$O(HWC_{i}C_{o}/N)$ \\ 2048} \\
      \hline
      \cellcolor{gray!25}{\#Aut. (Rot.)} & 0 & \makecell{$O(\lceil HW/N \rceil \log(N/HW)C_o)$ \\ 1280} & \makecell{$O(\sqrt{HWC_i C_o/N})$ \\ 10} \\
      \hline
      \cellcolor{gray!25}{\#Input Ct.} & \makecell{$O( HWC_{i}/N)$ \\ 8} & \makecell{$O( HWC_{i}/N)$ \\ 8} & \makecell{$O( HWC_{i}/N)$ \\ 8} \\
      \hline
      \cellcolor{gray!25}{\#Output Ct.} & \makecell{$O(\lceil HW/N \rceil C_{o})$ \\ 256} & \makecell{$O(HWC_o/N)$ \\ 8} & \makecell{$O(HWC_o/N)$ \\ 8} \\
      \hline
      \cellcolor{gray!25}{2PC moduli} & PO2 & PO2 & PO2 \\
      \hline\hline
      \cellcolor{gray!25}{\textbf{SIMD}} & \cellcolor{gray!25}{\textbf{Orion\cite{ebel2025orion}}} & \cellcolor{gray!25}{\textbf{Hyena\cite{singh2024hyena}}} & \cellcolor{gray!25}{\textbf{Gazelle\cite{juvekar_gazelle_2018}}} \\
      \hline\hline
      \cellcolor{gray!25}{\#CPMult} & \makecell{$O(C_{o}C_{i}HWf/N)$ \\ 18432} & \makecell{$O(HWC_{i}C_{o}f/N)$ \\ 38016} & \makecell{$O(C_{i}C_{o}HWf/N)$ \\ 36864} \\
      \hline
      \cellcolor{gray!25}{\#Aut. (Rot.)} & \makecell{$O(\sqrt{HWf/N}(C_i+C_o))$ \\ 328} & \makecell[c]{$O(f(C_{i}HW/N$\\$+C_{o}f/HW \log(fN/HW)))$ \\ 920} & \makecell{$O(C_{i} \cdot f)$ \\ 2288} \\
      \hline
      \cellcolor{gray!25}{\#Input Ct.} & \makecell{$O(C_{i}HW/N)$ \\ 8} & \makecell{$O(HWC_{i}/N)$ \\ 16} & \makecell{$O(C_{i}HW/N)$ \\ 16} \\
      \hline
      \cellcolor{gray!25}{\#Output Ct.} & \makecell{$O(C_{o}HW/N)$ \\ 8} & \makecell{$O(HWC_{o}/N)$ \\ 16} & \makecell{$O(C_{o}HW/N)$ \\ 16} \\
      \hline
      \cellcolor{gray!25}{2PC moduli} & Prime & Prime & Prime \\
      \hline
    \end{tabular}%
  }
  \label{tab: method comparison}%
\end{table}%

Based on these properties, we present two \textbf{key insights}, illustrated in Figure \ref{fig: 4-encoding} (f): \textbf{1)} We can apply the automorphism to the input ciphertext $\hat{x}$ and plaintext $\hat{w}$ beforehand rather than on the result $\hat{y}_i$ as there are less $\hat{x}$. For instance, in our example, there are only one input ciphertext $\hat{x}$, but four intermediate output ciphertexts \{$\hat{y}_1$,$\hat{y}_2$,$\hat{y}_3$,$\hat{y}_4$\}. \textbf{2)} BSGS algorithm can be applied if the automorphisms form a geometric progression.

Therefore, we reformulate the computation as shown in Figure \ref{fig: 4-encoding} (g). 
First, we observe that the cleaned ciphertext $\hat{y}''_i$ is obtained by computing $\hat{y}_i' = \hat{y}_i + \sigma(\hat{y}_i, N+1)$, followed by $\hat{y}_i'' = \hat{y}_i' + \sigma(\hat{y}_i', N/2+1)$. By substituting the $\hat{y}^{'}_i$ with $\hat{y}_i$, this sequence is equivalent to:
\[
\hat{y}_i'' = \hat{y}_i + \sigma(\hat{y}_i, N+1) + \sigma(\hat{y}_i, N/2+1) + \sigma(\sigma(\hat{y}_i, N+1), N/2+1)
\]
Let $\sigma$ denote the automorphism $\sigma(\cdot, N/2+1)$. By substituting this notation, the equation simplifies to a summation over the powers of $\sigma$:
\[
\hat{y}_i'' = \hat{y}_i + \sigma(\hat{y}_i) + \sigma^2(\hat{y}_i) + \sigma^3(\hat{y}_i) = \sum_{j=0}^3 \sigma^j(\hat{y}_i)
\]
Replacing the intermediate components $\hat{y}_i = \hat{x} \cdot \hat{w}_i$, the computation for a single ciphertext becomes:
\[
\hat{y}_i'' = \sum_{j=0}^3 \sigma^j(\hat{x}) \sigma^j(\hat{w}_i)
\]
We observe that the term $\sigma^j(\hat{x})$ remains the same across all $\hat{y}_i''$. Consequently, we can restructure the computation of the final output $\hat{y} = \sum_i \hat{y}''_i X^i$ by substituting $\hat{y}_i''$ into the equation:
\begin{multline*}
    \hat{y} = \sum_{i=0}^3 \sum_{j=0}^3 \sigma^j(\hat{x}) \sigma^j(\hat{w}_i) X^i \\ = \sum_{j=0}^3 \sigma^j(\hat{x}) \left( \sum_{i=0}^3 \sigma^j(\hat{w}_i) X^i \right) = \sum_j \sigma^j(\hat{x}) \hat{q}_j
\end{multline*}
The term in parentheses consists entirely of constant weights; therefore, we can precompute it and define $\hat{q}_j =  \sum_{i=0}^3 \sigma^j(\hat{w}_i) X^i$.
This reformulated dataflow is illustrated in Figure~\ref{fig: 4-encoding}~(g). 
Notably, this optimization reduces the number of required automorphisms in our example from eight to three.

The computation can be further optimized using the Baby-Step Giant-Step (BSGS) technique, as shown in Figure~\ref{fig: 4-encoding}~(h). Instead of computing the full set $\{\sigma(\hat{x}), \sigma^2(\hat{x}), \sigma^3(\hat{x})\}$, we compute only $\sigma(\hat{x})$. We then derive the final result by grouping terms and applying $\sigma^2$ as the "giant step":
\[
\hat{y} = \hat{x}\hat{q}_1 + \sigma(\hat{x})\hat{q}_2 + \sigma^2 \left( \hat{x}\sigma^{-2}(\hat{q}_3) + \sigma(\hat{x})\sigma^{-2}(\hat{q}_4) \right)
\]

This method reduces the required automorphisms from three to two. Generally, this process can be written as:
\begin{equation}
\label{eq: BSGS}
\begin{split}
    \mathbf{\hat{y}}_k &= \sum_{j=0}^{N_i-1}\sum_{i=0}^{2^k-1} \sigma^i(\hat{x}_j)\hat{q}_{ijk} \\
    &= \sum_{j=0}^{N_i-1}\sum_{i_1=0}^{2^{k_2}-1} \sigma^{i_1 \cdot 2^{k_1}} \left( \sum_{i_2=0}^{2^{k_1}-1} \sigma^{i_2}(\hat{x}_j)\hat{q}_{ijk} \right)
\end{split}
\end{equation}

where $N_i$ is the number of input ciphertexts (1 in this case) and $2^k$ is the number of channels contained in one ciphertexts (4 in this case). We provide a detailed comparison with other protocols regarding the number of HE operations in Table \ref{tab: method comparison}. Our technique saves $128 \times$ rotations compared to an implementation without the BSGS optimization. 

\section{Memory Aware Optimizations}
\label{sec: system optimization}

\subsection{Memory Bottlenecks of our protocol}

\begin{figure}[!tb]
    \centering
    \includegraphics[width=\linewidth]{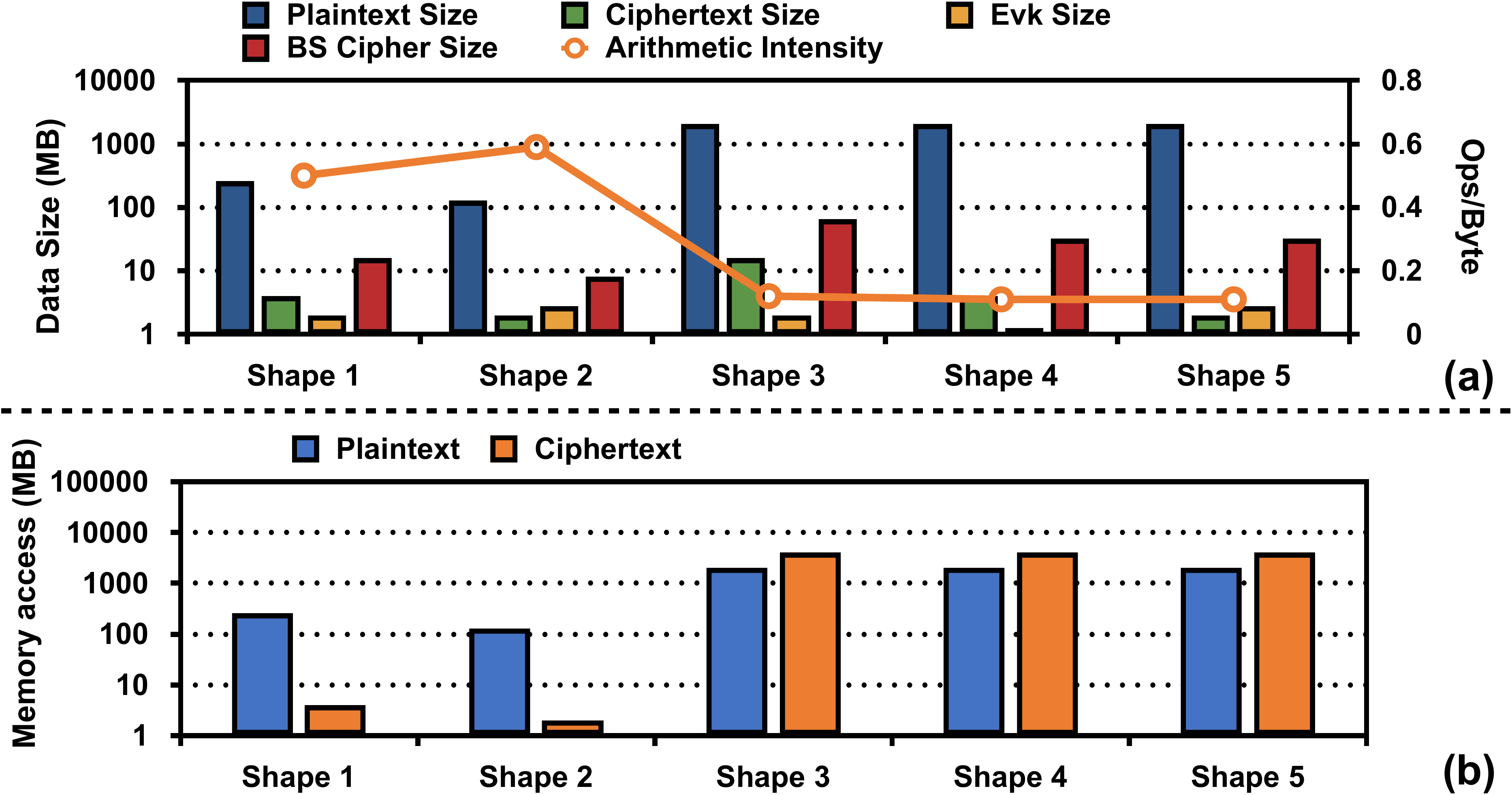}
    \caption{ (a) Data size and arithmetic intensity of several Resnet50 layers with Cheetah. "BS ciphertext" is the total ciphertext after baby step automorphisms. The baby step is chosen to minimize the total computation according to \cite{Halevi2018helib}. (b) Memory access of plaintext and ciphertexts.
    }
    \label{fig: 4-arithmetic}
\end{figure}

To identify the hardware performance bottlenecks of our proposed protocol, we profiled the data volume and memory access patterns of representative linear layers in ResNet-50, as shown in Figure~\ref{fig: 4-arithmetic}. Our analysis reveals three key findings: \textbf{1)} \textbf{Evaluation keys are not the primary bottleneck.} The memory footprint of evaluation keys is negligible for two reasons: the size of each key shrinks quadratically with its level $L$ (e.g., only 384KB for $L=1$), and the BSGS method drastically reduces the total number of keys required. Consequently, the dominant memory overhead stems from plaintext and ciphertext data movement. \textbf{2)} \textbf{Plaintexts create a high-volume data bottleneck.} As shown in Figure~\ref{fig: 4-arithmetic}(b), the large volume of plaintext data creates significant memory traffic. This is a direct consequence of our protocol's sparse encoding scheme (visualized in Figure~\ref{fig: 6-system}(a)), which requires a large number of weight plaintext polynomials. \textbf{3)} \textbf{Ciphertexts create a memory thrashing bottleneck.} The "baby step" phase of the BSGS algorithm generates a large working set of temporary ciphertexts that exceeds the on-chip scratchpad capacity. Although these ciphertexts are reused intensively during the subsequent "giant step" computations, their large collective size forces constant data swapping (thrashing) between the scratchpad and off-chip HBM. This results in the high volume of DRAM reads observed in Figure~\ref{fig: 4-arithmetic}(b).

As shown in Figure~\ref{fig: 4-arithmetic}, the arithmetic intensity of our protocol is merely 0.6 Ops/Byte, classifying the workload as heavily memory-bound due to the large data movement required for both plaintexts and ciphertexts. To resolve these two challenges, we introduce two hardware optimizations. To tackle the high volume of plaintext data, we propose \compression, a lightweight compression system that exploits inherent data sparsity to reduce memory traffic. To resolve the ciphertext thrashing problem, we co-design \dataflow, a reuse-centric dataflow that reorders computation to maximize on-chip data locality and reuse.

\subsection{\compression: Plaintext Compression System}

\begin{figure}[!tb]
    \centering
    \includegraphics[width=\linewidth]{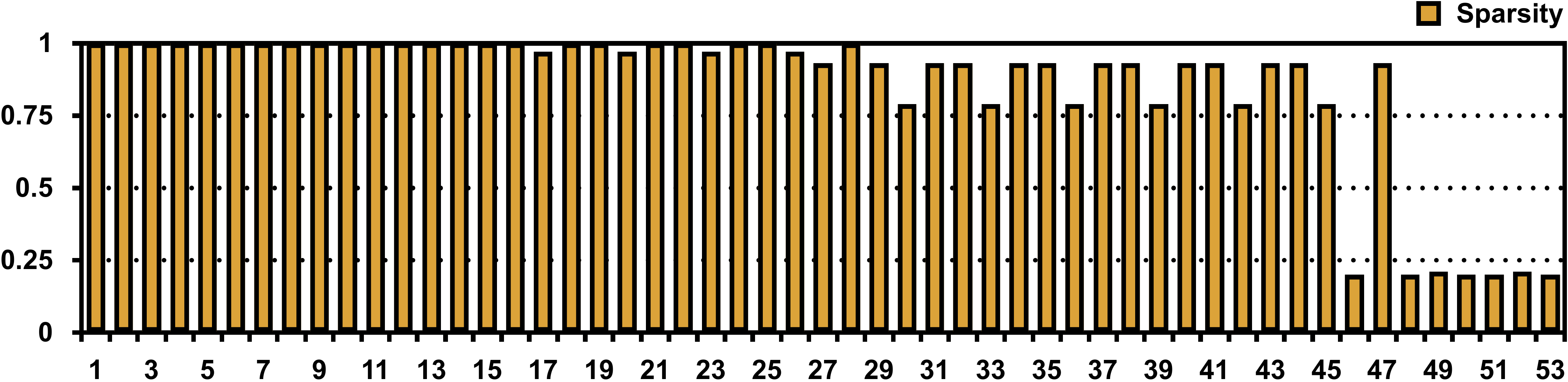}
    \caption{Sparsity in each layer of ResNet50 (higher means sparser).}
    \label{fig:resnet_coeff_sparsity}
\end{figure}

\begin{figure}[!tb]
    \centering
    \includegraphics[width=\linewidth]{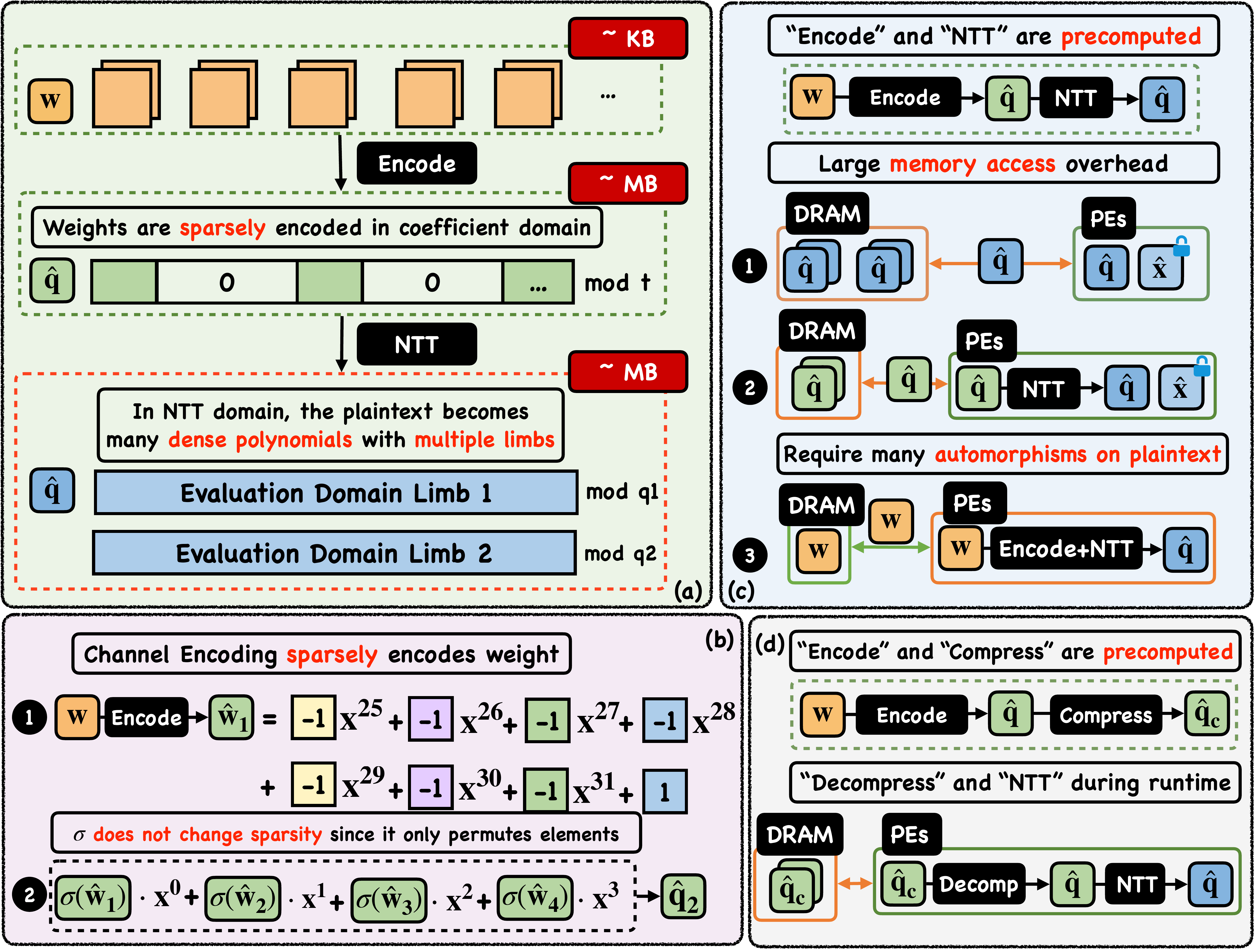}
    \caption{
    (a) The overall precomputation process.
(b) Our application's encoding process. A weight tensor is encoded into a \textbf{sparse} plaintext polynomial and then summed to derive $\mathbf{\hat{q}_i=\sum_j \sigma^i(\hat{w_j})x^j}$.
(c) A comparison of three plaintext management strategies: Method \textcircled{1} precomputes both encoding and NTT, as used in current libraries. Method \textcircled{2} precomputes the encoding and performs NTT on-chip. Method \textcircled{3} performs both encoding and NTT on-chip.
(d) Our proposed method: Leveraging the plaintext's sparsity, we precompute the encoding and compress the result. At runtime, the data is decompressed before the on-chip NTT is performed.
    }  
    \label{fig: 6-system}
\end{figure}

We begin by illustrating why weight plaintexts in our application is sparse. As shown in Figure \ref{fig: 6-system} (b), the plaintext encoding takes two steps. First, weight tensors are encoded into the plaintext with channel encoding, which is highly sparse (8 non-zero coefficients for $N=32$ in this example). Then we compute $\mathbf{\hat{q}_i=\sum_j \sigma^i(\hat{w_j})x^j}$ to get the plaintext we really use. Notably, the $\sigma$ operation does not change the sparsity; it only permutes the coefficients, and only the summation might slightly reduce sparsity. We have also empirically verified this by profiling ResNet50, which shows high sparsity in most layers as in Figure~\ref{fig:resnet_coeff_sparsity}.

Given this sparsity, two naive solutions could be considered as shown in Figure~\ref{fig: 6-system}~(c). The first is to store plaintexts in the plaintext ring $R_t$ in DRAM and perform the NTT on-chip (\textcircled{2}). This approach is insufficient, as it only halves the memory traffic, which still remains a significant bottleneck. The second approach is to store the original weights and perform both encoding and NTT on-chip (\textcircled{3}). This is also infeasible because it requires performing automorphisms on a vast number of plaintexts; while cheaper than ciphertext automorphisms, the large volume makes this prohibitively expensive.

Therefore, as shown in Figure \ref{fig: 6-system}~(d), we propose a third approach: directly storing the compressed plaintext in DRAM and decompressing it on-chip. This method avoids the high overhead of both large data transfers and on-chip plaintext automorphisms, offering a more effective solution to the memory bottleneck.

\begin{figure}[!tb]
    \centering
    \includegraphics[width=\linewidth]{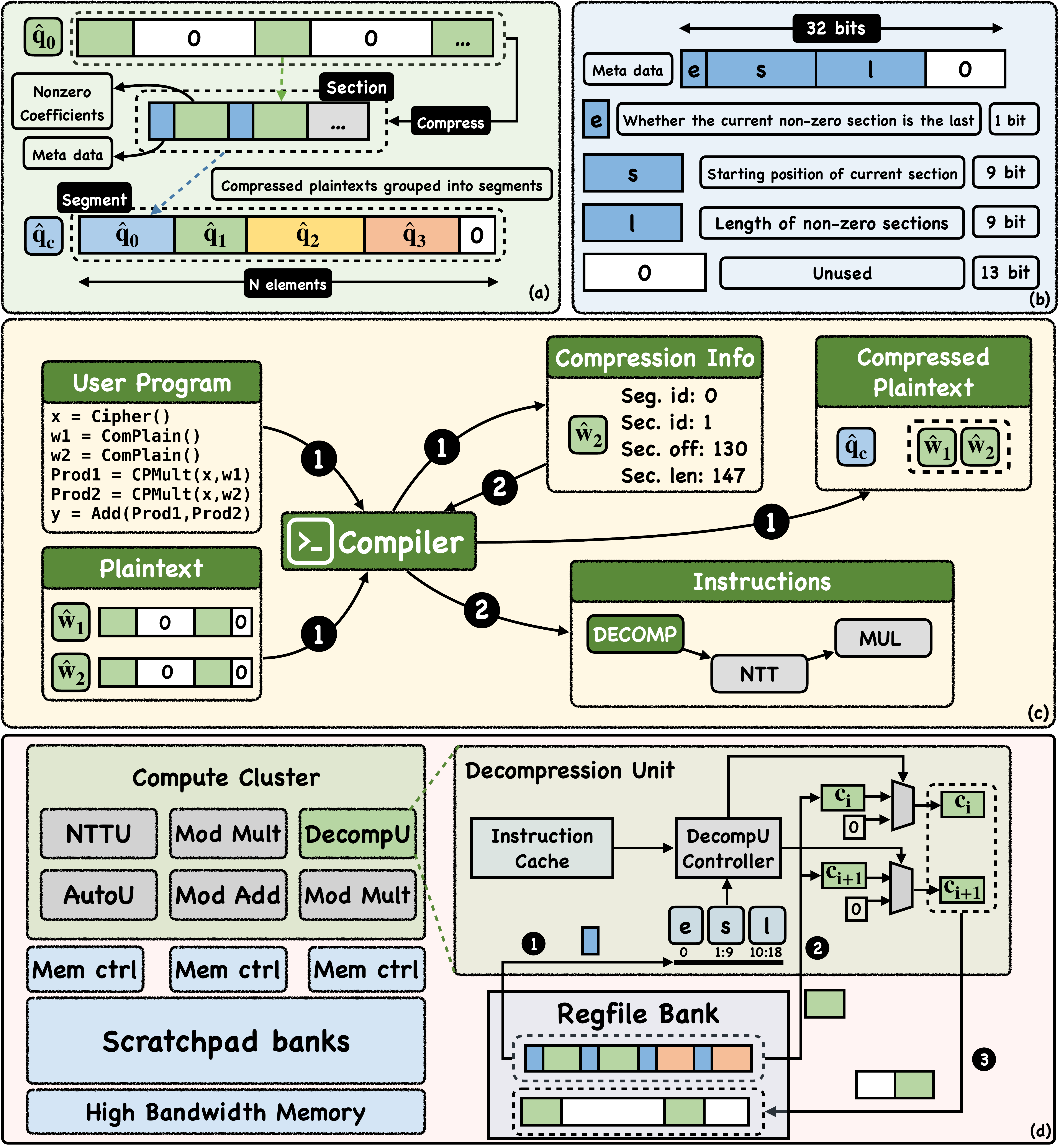}
    \caption{
    (a) The compressed plaintext data structure. (b) The metadata structure.
    (c) Modification to the compiler. The compiler needs to first take in the plaintext and the user program, then output the compression info and the compressed plaintext. Then, the compiler will generate the output using the compression info. (d) The hardware architecture of the decompression unit.}
    \label{fig: overall architecture}
\end{figure}

Figure~\ref{fig: overall architecture} illustrates our proposed plaintext compression system. We profile the plaintexts of our protocols and find that all plaintexts exhibit a sparse pattern in Figure~\ref{fig: overall architecture}~(a), which consists of a few dense blocks of non-zero coefficients separated by long runs of zeros. We believe this pattern comes from channel encoding, which encodes weight elements in several clusters, and the automorphisms in our protocol also preserve this structure, as most elements are kept or negated in place. The compression strategy is to discard these zero-runs and store only the non-zero data blocks. To enable reconstruction, each non-zero block is paired with metadata specifying its length and original starting position in the polynomial. Thus, a complete sparse polynomial is compressed into a compact representation consisting of a series of these metadata-data pairs.

Storing these variable-length non-zero sections sequentially in memory is problematic. Since existing memory systems manage data in fixed-size units of a full RNS polynomial, a direct layout would create memory alignment issues and complicate address calculation. To resolve this, we introduce a fixed-size container called a "segment," as shown in Figure~\ref{fig: overall architecture}~(a). A segment is constructed by packing multiple non-zero sections contiguously and zero-padding the remainder to match the full polynomial length, $N$. As detailed in Figure~\ref{fig: overall architecture}~(b), the metadata for each packed section has three components: its original starting position, its length, and an "end bit" (`e`) to flag if it is the final section of the original sparse polynomial.

Our compiler is modified to support this new data format, as outlined in Figure~\ref{fig: overall architecture}~(c). Programmers annotate plaintexts for compression by using a new `ComPlain()` data type. The compiler then processes these annotations in a new, initial pass (\textcircled{1}). In this pass, it packs the compressed plaintexts into fixed-size segments, grouping them in the order they appear in the program. For each plaintext, the compiler generates "compression information"—its location, specified by a segment index, a byte offset, and a length. This information is used by later compiler passes to generate the final machine code. To handle decompression at runtime, we introduce a new instruction, \textbf{Decomp Len, Off}. The compiler inserts this instruction before any operation (like NTT) that needs the full plaintext. The `Len` and `Off` operands tell the on-chip decompression unit exactly which portion of a segment to read and decompress.

To support our compression scheme, we integrate a dedicated decompression unit into each Compute Cluster as shown in Figure \ref{fig: overall architecture}~(d). The overall workflow is as follows: prior to a computation like NTT, a compressed plaintext segment is fetched into an on-chip scratchpad. The decompression unit then reads this segment and reconstructs the original, full-length sparse polynomial. The unit itself consists of a simple controller and a set of multiplexers. During decompression, the controller parses the metadata for each non-zero section to determine its length and its target offset in the final polynomial. It then writes the non-zero data to the correct locations in the register bank while filling the rest with zeros.

\begin{figure}[!tb]
    \centering
    \includegraphics[width=\linewidth]{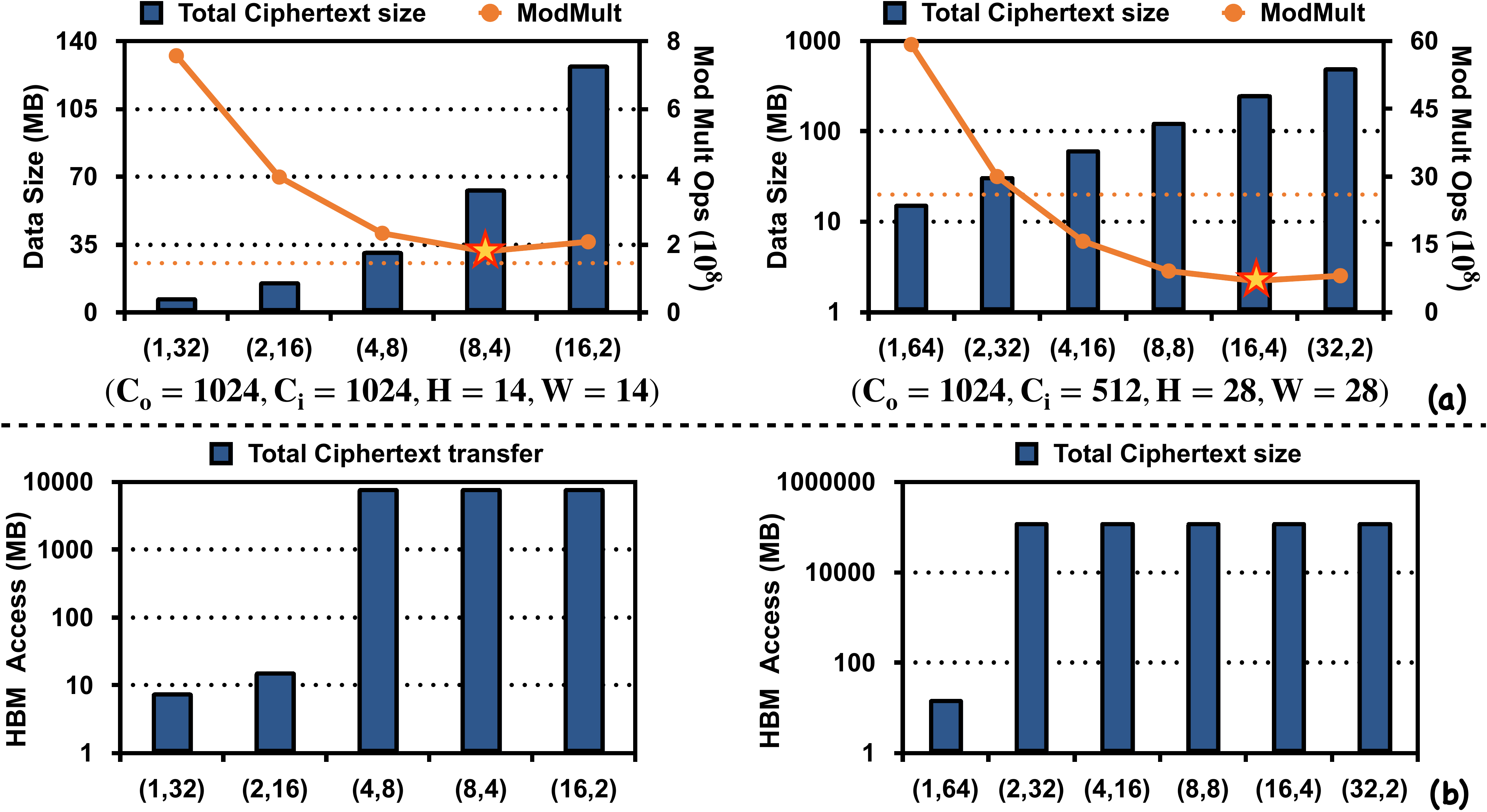}
    \caption{(a) The total computation and temporary ciphertext with varying baby steps. The orange dotted line represents the total on-chip SRAM size (25MB). \revB{The x-axis represents the baby step and giant step split.} (b) The temporary ciphertexts' memory access overhead. Those who suffer from the thrashing problem incur huge memory access.} 
    \label{fig:BSGS}
\end{figure}

\begin{figure}[!tb]
    \centering
    \includegraphics[width=\linewidth]{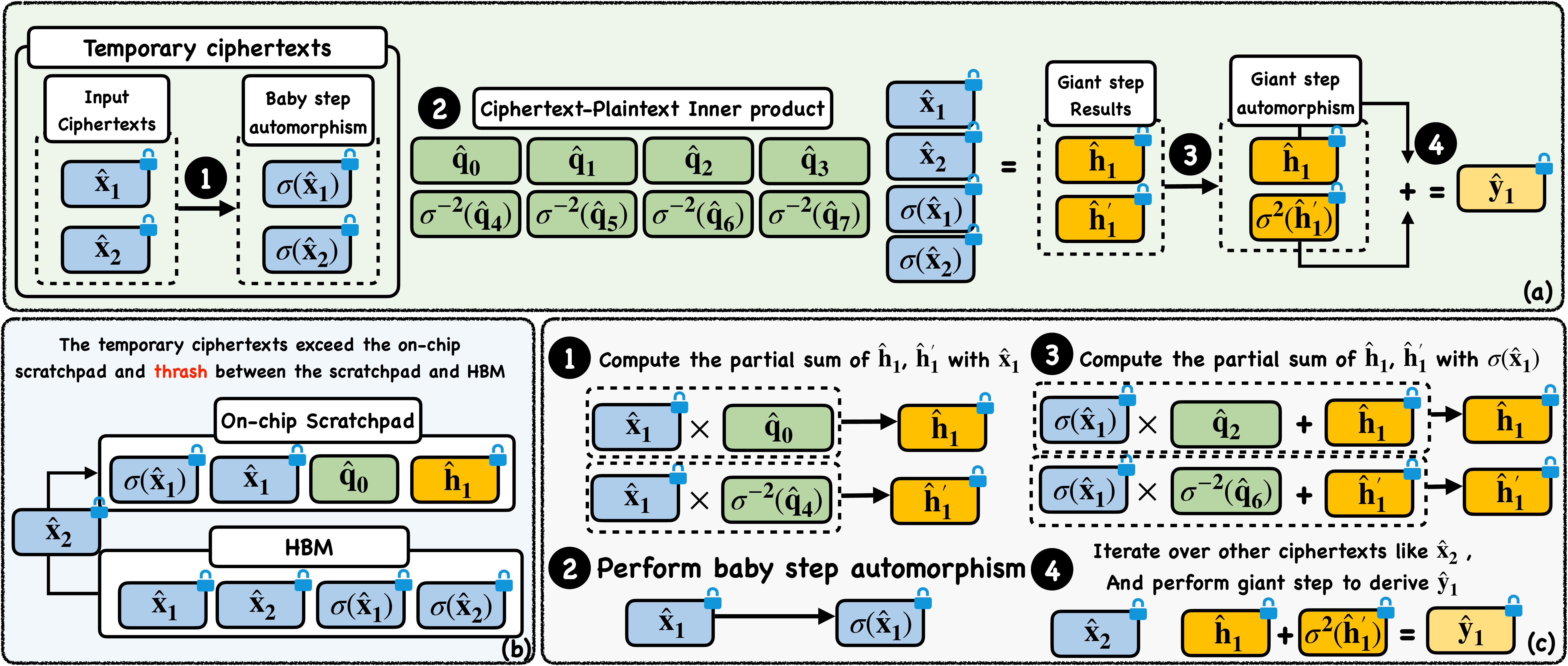}
    \caption{(a) An example of \method's protocol. (b) The temporary ciphertext thrashing problem is due to the limited scratchpad size. (c) Output stationary dataflow to minimize the ciphertext data movement.} 
    \label{fig: BSGS Diagram}
\end{figure}

\subsection{\dataflow: A Reuse-Centric Dataflow}
\label{sec: dataflow}


The computation of our protocol is formulated using BSGS in section \ref{sec: efficient conv}. A standard implementation, shown in Figure~\ref{fig: BSGS Diagram}(a), proceeds in four phases: \textcircled{1} Baby step automorphisms are applied to all inputs, \textcircled{2} inner products are computed, \textcircled{3} giant step automorphisms are applied to the intermediate results, and \textcircled{4} the partial results are aggregated. For all computations, we use the hoisting optimization for key switching, as the keys are small enough to reside entirely in the on-chip scratchpad (Figure~\ref{fig: 4-arithmetic}).

 The optimal split of $k$ into $k_1$ and $k_2$ presents a critical trade-off. A standard split ($k_1 \approx k_2$) is suboptimal for our application because the hoisted baby-step computations are cheaper to compute and are heavily reused across all $N_o$ outputs. Our analysis (Figure~\ref{fig:BSGS}(a)) reveals that the most computationally efficient configuration favors a large number of baby steps ($k_1 > k_2$).

Unfortunately, this computationally optimal strategy is memory-inefficient. A large $k_1$ generates a massive working set of $N_i \cdot 2^{k_1}$ temporary ciphertexts (e.g., $\{\sigma(\hat{x}), \dots, \sigma^{2^{k_1}-1}(\hat{x})\}$), which exceeds the on-chip scratchpad capacities. Consequently, these temporary results must be stored in off-chip HBM. Despite their high reuse potential, they must be repeatedly fetched, leading to severe memory thrashing between the scratchpad and HBM. The resulting memory traffic, shown in Figure~\ref{fig:BSGS}(b), is immense and completely negates the computational benefits. Prior work such as SHARP~\cite{jongmin23sharp} avoids this by using a small $k_1$, but this choice incurs a significant computational penalty (e.g., $4\times$ more computation than optimal).

To resolve this dilemma, we propose \dataflow, a reuse-centric dataflow that achieves both the computation and memory access efficiency. The core idea is to process one input ciphertext at a time, calculating its contribution to all output ciphertexts before moving to the next input. This avoids materializing the entire enormous set of temporary ciphertexts at once.

We illustrate this output-stationary dataflow using the example in Figure~\ref{fig: BSGS Diagram}(c). In step \textcircled{1}, A single input ciphertext, $\mathbf{\hat{x}_j}$, is loaded from HBM into the on-chip scratchpad. The accelerator uses it to update the partial sums for all $N_o \cdot 2^{k_2}$ outputs that depend on it. These partial sums remain resident on-chip. \textcircled{2} The next baby step automorphism (e.g., $\sigma^1(\hat{x}_j)$) is computed from the previous one. Its results are immediately used to again accumulate into the on-chip partial sums. This is repeated for all $2^{k_1}$ baby steps for the input $\mathbf{\hat{x}_j}$. \textcircled{3} After all baby steps for $\mathbf{\hat{x}_j}$ are completed, its data is discarded. The process repeats by loading the next input ciphertext, $\mathbf{\hat{x}_{j+1}}$.
\textcircled{4} Once all $N_i$ input ciphertexts have been processed, the complete partial sums in the scratchpad undergo the final giant step automorphisms and aggregation to produce the $N_o$ final outputs.

This dataflow relies on the scratchpad being large enough to hold all partial sums ($N_o \cdot 2^{k_2}$ ciphertexts), which is feasible as the computationally optimal split uses a small $k_2$. If the partial sums still exceed capacity, we simply tile the computation along the output dimension $N_o$. Crucially, \dataflow{} recomputes the baby-step ciphertexts on-the-fly rather than storing them. With hoisting, the cost of an automorphism recomputation is far less than the latency of a single HBM access, making this a highly effective trade-off.

\begin{figure*}[!htbp]
    \centering
    \includegraphics[width=\linewidth]{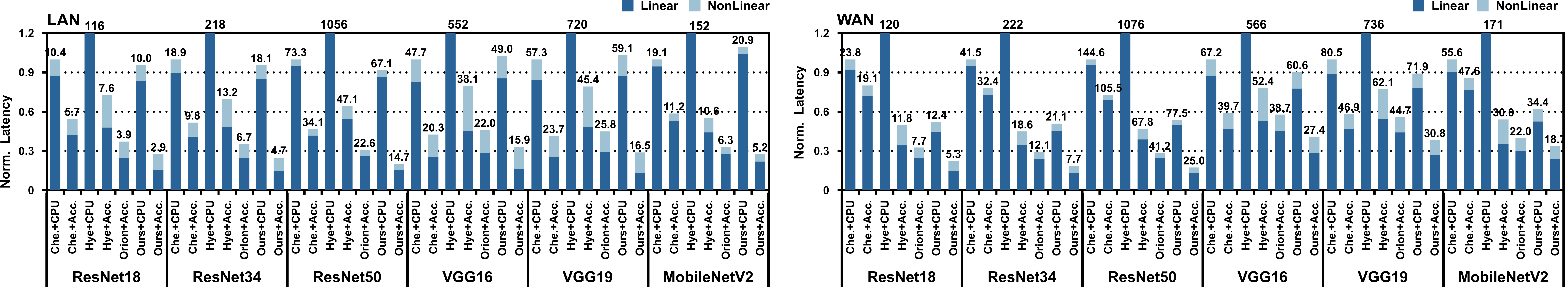}
    \caption{\revCC{End-to-end inference evaluation under LAN and WAN network conditions. ``+CPU'' and ``+Acc.'' denote the CPU-only and accelerator-augmented platforms.}}
    \label{fig:end2end}
\end{figure*}

\begin{figure*}[!htbp]
    \centering
    \includegraphics[width=\linewidth]{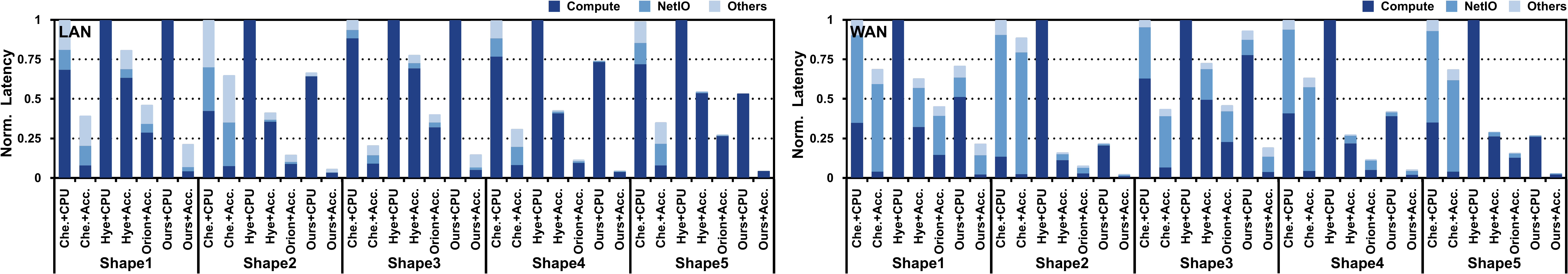}
    \caption{\revCC{Micro benchmark evaluation of linear layers.}}
    \label{fig: micro-benchmark}
\end{figure*}

\section{Evaluation}

\label{sec: experiments}

In this section, we evaluate \method, quantify our improvements over prior work, and demonstrate the effectiveness of our approach. We begin by presenting our evaluations across different CNN networks on ImageNet \cite{Jia2009imagenet} in Figure \ref{fig:end2end}. We then provide a detailed analysis of the benefits of our protocol and our memory-efficient optimization.

\subsection{Methodology}

\paragraph{Implementation}
\restored{We implement \method~on the LattiSense platform~\cite{lattisense}. Using the provided toolchain, \method's HE computation graph is defined in Python and compiled into accelerator instructions. The host program is implemented in C++ to perform encoding/decoding, encryption/decryption, network transmission of ciphertexts, and invocation of the accelerator. To support our memory-efficient contributions, we extend the compiler to generate instructions for the decompression unit and to optimize the Baby-Step Giant-Step (BSGS) dataflow. The decompression unit is implemented in Verilog. For the nonlinear layers, we integrate the implementations of CrypTFlow2 and Cheetah~\cite{rathee2020cryptflow2,huang2022cheetah} with the EMP toolkit~\cite{emp_toolkit} and the EzPC framework~\cite{chandran2019ezpc} in C++. We use the communication-efficient Vector-OLE-based Oblivious Transfer (VOLE-OT) protocol~\cite{Roy2022vole,yang2020ferret} across all experiments for a fair comparison.}
\paragraph{Experimental Setup} We evaluate \method~ on an Intel(R) Xeon(R) Gold 6226R CPU @ 2.90 GHz with 256 GB of RAM and use 16 threads. The FPGA accelerator is deployed on the AMD Alveo U55C Accelerator Card and interacts with the host CPU through PCIe. The network condition is simulated through Linux Traffic Control. The bandwidth is set to 384MBps for \texttt{LAN} and 44MBps for \texttt{WAN}, with a round-trip latency of 0.3ms for \texttt{LAN} and 40ms for \texttt{WAN}. 
\paragraph{Baseline} The baseline for our protocol is two HE-MPC frameworks, a coefficient encoding protocol Cheetah \cite{huang2022cheetah} and SIMD encoding Hyena \revB{\mbox{\cite{singh2024hyena}}} for convolution operations. \revCC{We further adapt the SIMD-based FHE protocol Orion~\mbox{\cite{ebel2025orion}} to the hybrid HE-MPC setting as an additional encoding baseline.} Since the main contribution of our work is not the accelerator itself, we do not compare it with other accelerators and instead treat it as a general platform.
\paragraph{Parameters}
We target 128-bit security for both HE and OT protocols. For HE, the multiplication depth is 1, we use the ciphertext modulus $q \approx 2^{64}$, \revA{special prime $p \approx 2^{32}$}, the plaintext modulus $t = 2^{21}$, and the polynomial degree $N = 8192$. For VOLE-OT, we select the security parameter $\lambda = 128$. The same set of cryptographic parameters is used across all experiments, with $t \approx 2^{21}$ chosen for the Hyena baseline, as it employs SIMD encoding.
\paragraph{Benchmarks}
\restored{For benchmarks, we use the ImageNet dataset~\cite{Jia2009imagenet}, whose image size is $20$--$50\times$ larger than the previously used CIFAR-10/100~\cite{krizhevsky2009cifar} and TinyImageNet~\cite{krizhevsky2009tinyimagenet} datasets. We evaluate end-to-end latency on ResNet~\cite{he2015resnet}, VGG~\cite{simonyan2015vgg}, and MobileNetV2~\cite{sandler2019mobilenetv2}.}

%

\subsection{End-to-End Performance Evaluation}

Figure~\ref{fig:end2end} presents our end-to-end latency evaluation on both CPU and accelerator platforms with \revCC{four} encodings, with all results normalized to the Cheetah+CPU baseline. Our method consistently outperforms the baselines across all configurations. Compared to CPU-only implementations, \method~ reduces latency to just $0.02\times$–$0.3\times$ in a \texttt{LAN} setting and $0.03\times$–$0.56\times$ in a \texttt{WAN} setting. When compared against scenarios with accelerators, \method~ achieved normalized latencies of $0.4\times$–$0.78\times$ (\texttt{LAN}) and $0.29\times$–$0.65\times$ (\texttt{WAN}). These substantial performance gains come from two factors: our protocol's communication efficiency and our accelerator's codesigned compression unit and dataflow, which reduces computation time. The difference for Hyena from its original publication comes from our larger polynomial degree ($N=8192$ vs. $N=1024$), which adds the automorphisms its SIMD encoding needs to aggregate partial sums. \revCC{Compared to Orion, \mbox{\method{}} is \mbox{$1.4$}--\mbox{$1.6\times$} faster under \mbox{\texttt{WAN}} and \mbox{$1.3$}--\mbox{$1.5\times$} under \mbox{\texttt{LAN}}, since its coefficient encoding incurs far fewer HE operations and less communication than Orion's SIMD encoding.}

Figure~\ref{fig: micro-benchmark} provides a detailed latency breakdown for representative ResNet50 layers to dissect the performance of \method, with all results normalized to the Cheetah+CPU baseline. Compared to Cheetah, our protocol design delivers a significant reduction in network communication latency in both network conditions with a small amount of extra computation. Furthermore, our method reduces CPU-side overhead for en/decryption and en/decoding (labeled "Others") by generating fewer output ciphertexts for the client to process.
For Hyena, while its network latency is also small compared to Cheetah, its overall performance is bottlenecked by a computational cost that is over $10\times$ higher, highlighting the effectiveness of our communication and computation efficient protocol. \revCC{Orion also keeps the network latency low through its compact output packing, but its per-layer HE compute stays higher than \mbox{\method}'s, so its latency remains above ours across all the representative layers.}

\subsection{Evaluation of \method's Protocol}

\begin{figure*}[!tb]
    \centering
    \includegraphics[width=\linewidth]{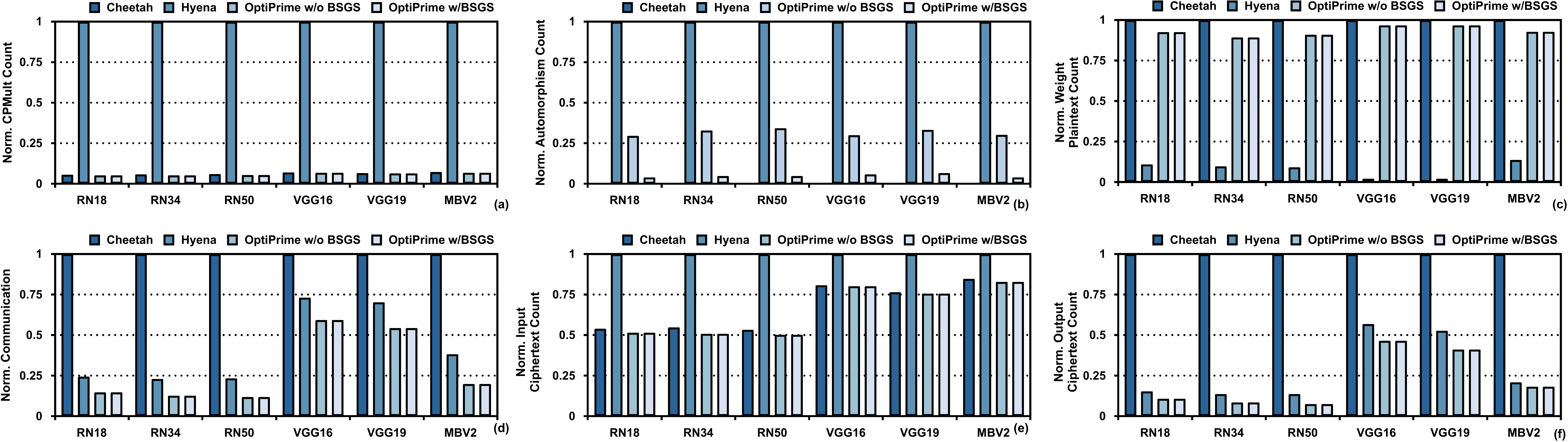}
    \caption{Operations Count and Communication Comparison against Baselines. "Norm." represents normalized. "CPMult" represents ciphertext-plaintext multiplication.}
    \label{fig: Protocol Evaluation}
\end{figure*}

Figure~\ref{fig: Protocol Evaluation}~(d--f) presents an evaluation of our protocol's communication efficiency against other baselines. Compared to Cheetah, our protocol reduces network communication to $0.11\times$-$0.59\times$ and the number of output ciphertexts to $0.07\times$-$0.46\times$. Compared to Hyena, the number of output ciphertexts is comparable, as the SIMD encoding with masking can also eliminate dummy slots~\cite{pang2023bolt}. \restored{Finally, the number of input ciphertexts remains similar across all three schemes because the input data is typically densely encoded. The difference mainly comes from their padding strategies. For example, Hyena pads each kernel to a power of two to accumulate the partial results of one convolution kernel. This difference has little effect on total communication because the output ciphertexts contribute the dominant overhead.}

We also evaluate computational efficiency by comparing the required number of ciphertext-plaintext multiplications (CPMult) and automorphisms. As detailed in Figure~\ref{fig: Protocol Evaluation}~(a) and (b), our protocol requires only $0.05\times$-$0.07\times$ the CPMults and $0.04\times$-$0.07\times$ the automorphisms of the Hyena protocol. This twofold advantage is attributed to the Baby-Step Giant-Step (BSGS) technique and the use of a coefficient-based encoding scheme, which is inherently suited for convolution. Notably, though the BSGS changes the content of each weight plaintext, it does not increase communication overhead, the number of CPMults, or the number of weight plaintexts, and is purely a free lunch. Compared to an implementation of \method~ without BSGS, our protocol requires just $0.13\times$-$0.19\times$ the automorphisms, further highlighting the efficacy of our approach. Although Cheetah requires no automorphisms, it generates a large volume of output ciphertexts, a trade-off that is unfavorable in scenarios with an accelerator where communication is a primary bottleneck.

Figure~\ref{fig: Protocol Evaluation}~(c) also compares the number of weight plaintexts required by each protocol. Our approach uses a larger number of plaintexts due to a sparse encoding of weights, which increases memory traffic and are optimized through hardware codesign. In contrast, while the Hyena protocol reduces the plaintext count, it does so at the cost of more computationally intensive operations (e.g., automorphisms) and remains inefficient over wireless networks.

Consequently, although our encoding incurs a minor increase in automorphisms, it achieves a substantial reduction in communication overhead. This trade-off proves highly favorable, as with a hardware accelerator, this small computational cost has a negligible impact on the overall end-to-end latency but contributes to larger communication reduction, which greatly impacts the overall latency and is hard to accelerate.

\subsection{Evaluation of Memory Optimization}

\begin{figure*}[!tb]
    \centering
    \includegraphics[width=\linewidth]{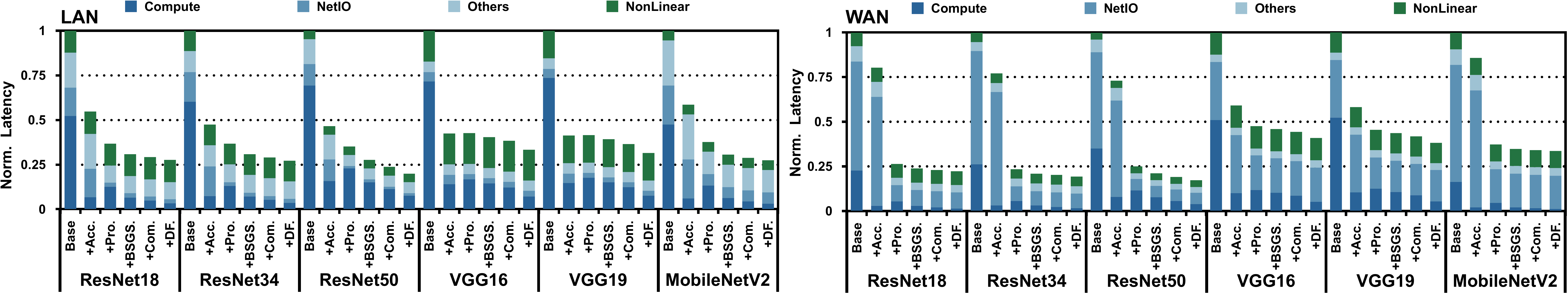}
    \caption{\revB{Ablation from the Cheetah+CPU baseline, adding one optimization at a time, with latency normalized to the baseline. ``+Acc.'' adds the HE accelerator; ``+Pro.'' adds our channel-encoding protocol; ``+BSGS'' adds the BSGS automorphism reduction; ``+Com.'' adds plaintext compression; ``+DF.'' adds the ciphertext dataflow.}}
    \label{fig:abaltion}
\end{figure*}

\begin{table}[!tb]
    \caption{Resource utilization for the decompression unit.}
    \label{tab:resources}
    \centering
    \begin{tabular}{c|c|c|c}
    \toprule
    LUT    & Reg & BRAM & DSP  \\ 
    \midrule
    0.8K & 0.8K  & 0  & 0 \\ 
    \bottomrule
    \end{tabular}
\end{table}

The decompression unit was implemented in RTL and integrated with the surrounding logic in collaboration with the LattiSense team. As detailed in Table~\ref{tab:resources}, the implementation utilizes 0.8k LUTs and 0.8K registers, constituting a mere 0.7\% of the total available on-chip resources the AMD U55C Accelerator Card (1304k LUTs and 2607k registers).

\begin{figure}[!tb]
    \centering
    \includegraphics[width=\linewidth]{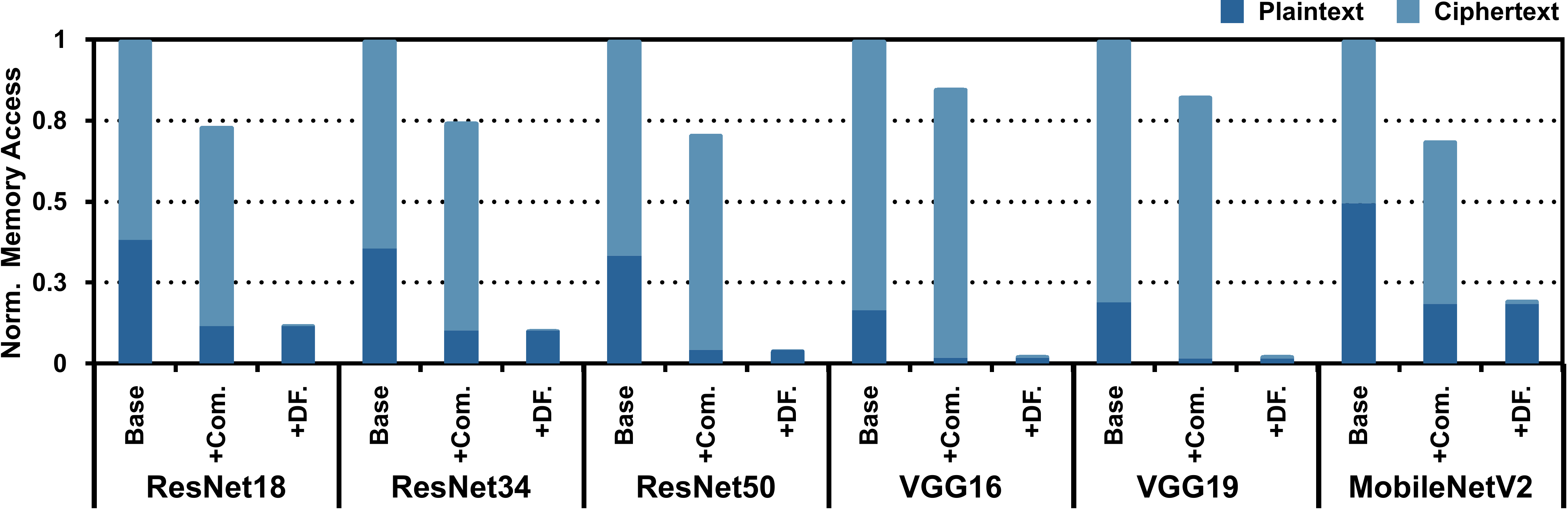}
    \caption{Memory access reduction through the plaintext compression technique and ciphertext dataflow. "Com." represents the plaintext compression technique, and "DF." represents the ciphertext dataflow technique.}
    \label{fig: Memory Opt}
\end{figure}

Figure~\ref{fig: Memory Opt} illustrates the reduction in memory access achieved by \compression~ and \dataflow~ across several CNN architectures. The baseline configuration stores all plaintexts in the evaluation domain and employs the computation-efficient 'baby step' selection of the BSGS algorithm. First, applying our plaintext compression technique (denoted ``+Com.'') reduces memory access to $0.7\times$-$0.85\times$ that of the baseline, underscoring the effectiveness of plaintext compression. This also shows that for our encoding, the sparsity of the weight plaintext is prevalent, as all networks benefit from the exploitation of sparsity. Second, by incorporating \dataflow (``+DF.''), we further reduce the memory access overhead to just $0.03\times$-$0.21\times$ of the baseline. This large reduction is attributed to the high data reusability enabled by our dataflow, which effectively mitigates memory thrashing.

\subsection{Ablation study}

\revB{Figure~\mbox{\ref{fig:abaltion}} reports an ablation that starts from the Cheetah+CPU baseline and adds one optimization at a time, with the latency normalized to the baseline. ``+Acc.'' offloads the HE computation to the accelerator and shrinks the compute part, bringing a \mbox{$1.7$}--\mbox{$2.4\times$} (LAN) / \mbox{$1.2$}--\mbox{$1.7\times$} (WAN) end-to-end speedup, after which the latency is dominated by the network communication. ``+Pro.'' applies our channel encoding, which reduces the network communication; this is the single largest step on WAN (up to \mbox{$3.3\times$} speedup); on LAN the gain is \mbox{$1.0$}--\mbox{$1.6\times$}, as the extra compute partly offsets the communication saving. ``+BSGS'' then eliminates that extra compute by reducing the automorphisms of the channel encoding, bringing a \mbox{$1.06$}--\mbox{$1.28\times$} (LAN) / \mbox{$1.04$}--\mbox{$1.18\times$} (WAN) speedup, largest on compute-heavy networks such as ResNet-50. Finally, ``+Com.'' reduces the plaintext memory access for a further end-to-end speedup of \mbox{$1.06$}--\mbox{$1.17\times$} (LAN) / \mbox{$1.02$}--\mbox{$1.10\times$} (WAN), and ``+DF.'' reduces the ciphertext memory access for a \mbox{$1.05$}--\mbox{$1.18\times$} (LAN) / \mbox{$1.01$}--\mbox{$1.11\times$} (WAN) latency reduction.}

\begin{figure}[!tb]
    \centering
    \includegraphics[width=\linewidth]{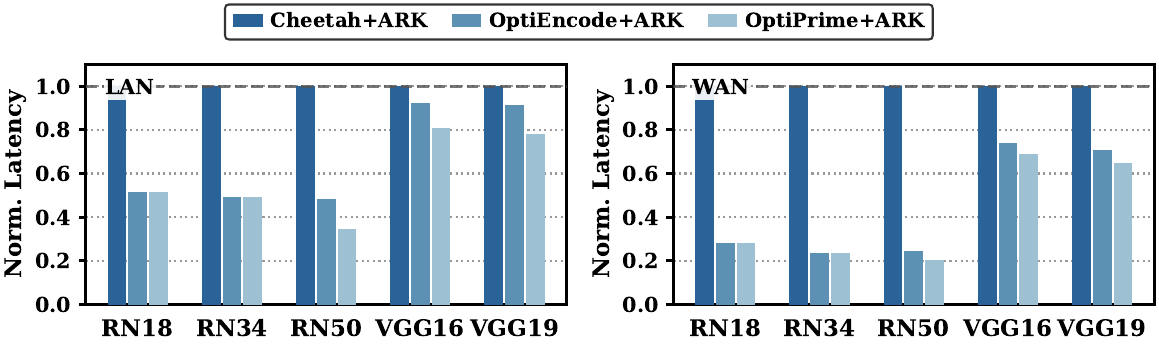}
    \caption{Normalized end-to-end latency evaluation when integrated with ARK \cite{jongmin22ark}.}
    \label{fig:e2ebreakdown}
\end{figure}
\subsection{Generality of \method{}}
\textbf{Generality to other accelerators.} We evaluate \method~on ARK~\cite{jongmin22ark} to demonstrate the generality of our work (Figure~\ref{fig:e2ebreakdown}). \restored{Because neither ARK hardware nor its simulator and compiler are publicly available, we construct a cycle-accurate simulator from ARK's published architectural parameters. We augment the modeled accelerator with our decompression unit and extend our compiler to emit ARK-compatible instructions, including the \texttt{Decomp} instruction, and to schedule the dataflow under ARK's 512\,MB scratchpad.} Compared to the \texttt{Cheetah+ARK} baseline, applying our protocol (\texttt{OptiEncode+ARK}) reduces latency to $0.48\times$--$0.51\times$ for ResNets and $0.91\times$--$0.92\times$ for VGGs under \texttt{LAN}, and $0.24\times$--$0.28\times$ / $0.71\times$--$0.74\times$ under \texttt{WAN}. With all optimizations (\texttt{OptiPrime+ARK}), latency further decreases to $0.35\times$--$0.51\times$ (\texttt{LAN}) and $0.21\times$--$0.28\times$ (\texttt{WAN}) for ResNets, and $0.78\times$--$0.81\times$ / $0.65\times$--$0.69\times$ for VGGs.

\begin{figure}[!t]
    \centering
    \includegraphics[width=\linewidth]{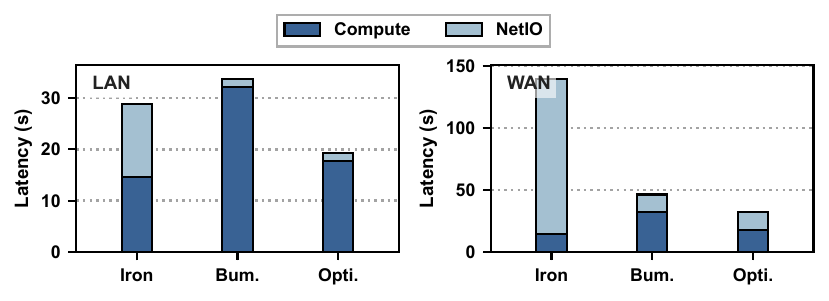}
    \caption{\revCC{Whole-network linear-layer latency on BERT-base for Iron, BumbleBee (Bum.), and \mbox{\protocol{}} (Opti.).}}
    \label{fig:transformer}
\end{figure}

\revCC{\textbf{Generality to Transformer models.} We evaluate \mbox{\method{}} on BERT-base against the prior-art protocols Iron~\mbox{\cite{hao2022iron}} and BumbleBee~\mbox{\cite{lu2023bumblebee}} to demonstrate its applicability to Transformer models. We implement each matrix multiplication as a $1\times1$ convolution. Under our LAN/WAN settings (Figure~\mbox{\ref{fig:transformer}}), \mbox{\method{}} is more efficient in both communication and computation, reducing the whole-network linear-layer latency to $0.67\times$/$0.23\times$ (LAN/WAN) that of Iron and $0.57\times$/$0.69\times$ that of BumbleBee.}

\begin{figure}[!t]
    \centering
    \includegraphics[width=\linewidth]{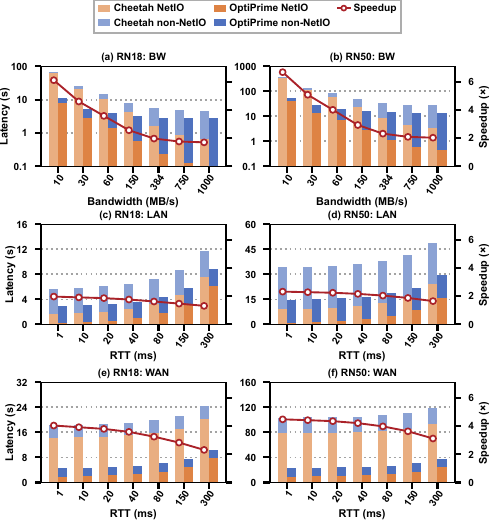}
    \vspace{-6pt}
    \caption{Network sensitivity for ResNet-18 (left) and ResNet-50 (right): bandwidth at 1\,ms RTT (top), RTT at LAN bandwidth (middle), and RTT at WAN bandwidth (bottom). Bars show NetIO/non-NetIO latency for Cheetah and \method{}; lines show speedup.}
    \label{fig:netsens}
\end{figure}

\begin{figure}[!t]
    \centering
    \includegraphics[width=\linewidth]{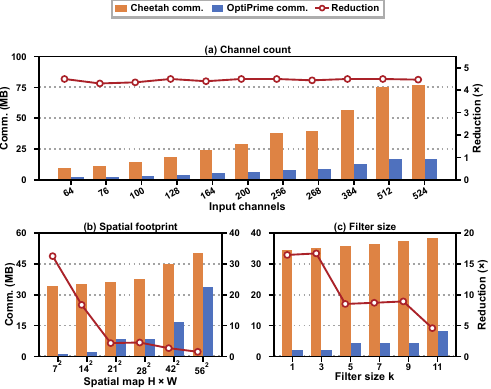}
    \vspace{-6pt}
    \caption{\revB{Performance of \mbox{\method{}} under nonideal (a) input-channel count, (b) spatial map size \mbox{$H\times W$}, and (c) filter size \mbox{$k$}, with the other dimensions fixed.}}
    \label{fig:tiling}
\end{figure}

\begin{figure}[!t]
    \centering
    \includegraphics[width=\linewidth]{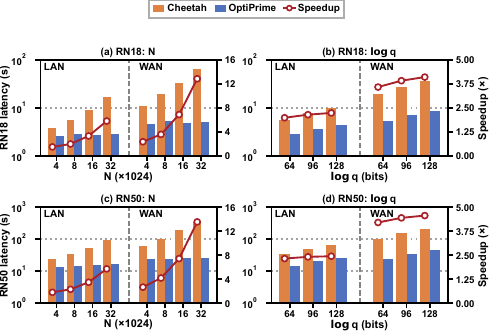}
    \vspace{-6pt}
    \caption{HE-parameter sensitivity on ResNet-18 (a--b) and ResNet-50 (c--d) under different polynomial degrees $N$ and ciphertext moduli $q$.}
    \label{fig:heparam}
\end{figure}

\subsection{Inference Accuracy}
\begin{table}[t]
\caption{\revCC{W8A8 ImageNet top-1 accuracy (\%) and measured convolution accumulator bit-width.}}
\label{tab: ptq-accuracy}
\centering
\resizebox{\linewidth}{!}{
\begin{tabular}{l|c|ccc|c}
\toprule
Model & FP32 & Cheetah & Hyena & \textbf{\method} & Accum.\ (bits) \\
\midrule
ResNet-18   & 69.76 & 69.51 & 69.51 & \textbf{69.51} & 18.5 \\
ResNet-34   & 73.30 & 73.04 & 73.04 & \textbf{73.04} & 18.8 \\
ResNet-50   & 76.14 & 75.93 & 75.93 & \textbf{75.93} & 19.1 \\
VGG-16      & 71.58 & 71.44 & 71.44 & \textbf{71.44} & 19.1 \\
VGG-19      & 72.39 & 72.33 & 72.33 & \textbf{72.33} & 19.3 \\
MobileNetV2 & 71.87 & 69.71 & 69.71 & \textbf{69.71} & 17.6 \\
\bottomrule
\end{tabular}
}
\end{table}

\revCC{We measure top-1 ImageNet accuracy under W8A8 quantization on pre-trained Torchvision models~\mbox{\cite{torchvision2016}}. Since \mbox{\method{}} alters only the intermediate computation and keeps the result exact, its accuracy matches prior protocols, with only negligible quantization loss versus FP32 (Table~\mbox{\ref{tab: ptq-accuracy}}).} \revD{The accumulator bit-width peaks at \mbox{$19.3$} bits, well below our plaintext modulus \mbox{$t=2^{21}$}, confirming the parameter set is valid.}

\subsection{Sensitivity Study}

\noindent \textbf{Network.} \restored{We evaluate ResNet-18 and ResNet-50 over bandwidths from 10\,MB/s to 1\,GB/s at a fixed 1\,ms RTT, and over RTTs from 1\,ms to 300\,ms at both LAN and WAN bandwidths (Figure~\ref{fig:netsens}). Across this bandwidth range, \mbox{\method{}} consistently outperforms Cheetah, achieving $1.71$--$6.12\times$ speedup on ResNet-18 and $2.04$--$6.69\times$ on ResNet-50. Even at 1\,GB/s, \mbox{\method{}} retains $1.71\times$/$2.04\times$ speedup, as its accelerator-side optimizations continue to reduce the non-network latency.}

\noindent \textbf{HE parameters} \revA{On ResNet-18, we sweep the polynomial degree \mbox{$N$} (at a fixed \mbox{$\log q\!=\!64$}) and the ciphertext modulus \mbox{$\log q$} (at a fixed \mbox{$N\!=\!8192$}) under LAN and WAN in Figure~\mbox{\ref{fig:heparam}}. The speedup over Cheetah holds at every \mbox{$N$} and even grows with it, on WAN from \mbox{$2.4\times$} at \mbox{$N\!=\!4096$} to \mbox{$12.9\times$} at \mbox{$N\!=\!32768$}, since our communication is invariant to \mbox{$N$} while Cheetah's grows.} \restored{ResNet-50 exhibits the same trend: its WAN speedup rises from $2.7\times$ to $13.6\times$ over the same $N$ sweep.} \revD{The speedup also holds as \mbox{$\log q$} grows from \mbox{$64$} to \mbox{$128$} bits, so the effectiveness of \mbox{\method} does not depend on the specific parameters \mbox{$(N,q,t)$}.} \restored{For ResNet-50 on WAN, it increases from $4.2\times$ to $4.6\times$ over this modulus range.}

\noindent \textbf{Layer shape.} \revB{Starting from a representative convolution layer, we vary one shape dimension at a time and report the communication of the transmitted ciphertexts and its reduction over Cheetah in Figure~\mbox{\ref{fig:tiling}}, to confirm that \mbox{\protocol} performs well when handling shapes that do not fit \mbox{$N$}. Sweeping the channel count \mbox{$C_i\!=\!C_o$} (at \mbox{$H\!=\!W\!=\!28$}, \mbox{$k\!=\!3$}), the reduction stays constant at \mbox{$4.5\times$}, as expected, since a non-ideal channel count only wastes space in the last ciphertext. Sweeping the spatial map \mbox{$H\!\times\!W$} (at \mbox{$C_i\!=\!C_o\!=\!256$}, \mbox{$k\!=\!3$}), it falls from \mbox{$32\times$} for small maps to \mbox{$\sim\!1.5\times$} for large maps; this does not mean our protocol degrades, but rather that a larger spatial map fills more useful slots in Cheetah's ciphertexts and leaves Cheetah less to waste, so the ratio shrinks because Cheetah improves. Sweeping the filter size \mbox{$k$} (at \mbox{$C_i\!=\!C_o\!=\!256$}, \mbox{$H\!=\!W\!=\!14$}), it drops from \mbox{$16.7\times$} (\mbox{$k\!\le\!3$}) to \mbox{$4.6\times$} (\mbox{$k\!=\!11$}) for the same reason. In every case, \mbox{\method{}} requires far less communication than Cheetah, demonstrating its robustness to non-ideal layer shapes.}

\section{Conclusion}
\label{sec: conclusion}
We propose \method, a protocol-hardware co-optimization framework for the hybrid HE-MPC setting. To address the wireless network communication bottleneck exposed by accelerators, we introduce a communication-efficient protocol based on channel encoding. We further deploy plaintext compression and a ciphertext-friendly dataflow to mitigate memory access overhead. Together, \method~ outperforms the Cheetah baseline by up to $5.7\times$ on CPUs and $4.2\times$ with an accelerator.

\section*{Acknowledgment}
This work was supported in part by NSFC under Grant 92464104, Grant 62495102,
and Grant 62341407, in part by the National Key Research and Development Program
under Grant 2024YFB4505004, in part by Beijing Advanced Innovation Center for
Future Blockchain and Privacy Computing, in part by 111 Project under Grant
B18001.

\bibliographystyle{IEEEtran}
\bibliography{reference/ref}

\end{document}